\documentclass[]{pasj02} 
\usepackage{natbib} 
\usepackage[utf8]{inputenc}
\newcommand{\Ism}{I_{\mathrm{sm}}}
\newcommand{\Isp}{I_{\mathrm{sp}}}
\newcommand{\Ismpix}[1]{I_{\mathrm{sm},#1}}
\newcommand{\Isppix}[1]{I_{\mathrm{sp},#1}}
\newcommand{\alphasm}{\alpha_{\mathrm{sm}}}
\newcommand{\alphasp}{\alpha_{\mathrm{sp}}}
\newcommand{\alphabl}{\alpha_{\mathrm{bl}}}

\jyear{2024}
\Received{}
\Accepted{}

\begin{document} 

\title{A New PSF Deconvolution Algorithm: Simultaneous Spatial Resolution Enhancement and Point Source Removal for Morphological Analysis of AGN Host Galaxies}

\author{
 Ren \textsc{Kawase},\altaffilmark{1}\altemailmark\email{m3255300095@std.kitami-it.ac.jp} 
 Takatoshi \textsc{Shibuya},\altaffilmark{1}\altemailmark\email{tshibuya@mail.kitami-it.ac.jp}
 and 
 Kazunori \textsc{Matsuda}\altaffilmark{1}\orcid{0009-0005-3887-0030}
}
\altaffiltext{1}{Kitami Institute of Technology, 165 Koen-cho, Kitami, Hokkaido 090-8507, Japan}



\maketitle

\begin{abstract}
We propose a new point-spread function (PSF) deconvolution algorithm for images of galaxies hosting an active galactic nucleus (AGN), designed to simultaneously enhance the spatial resolution of the host galaxy and remove the bright central point source. In this algorithm, an intrinsic image is reconstructed by decomposing an observed image into two components: an image $\Ism$ of an extended component (i.e., a host galaxy) and an image $\Isp$ of a point-source component (i.e., an AGN). During image reconstruction, three constraints are imposed: (1) a smooth constraint on the image $\Ism$, which spatially smooths the host-galaxy structures; (2) a sparse constraint on the image $\Isp$, which localizes the point source to a small number of pixels; and (3) a new constraint, the point-source balance constraint, based on the pixel-wise product $\Ism \times \Isp$, which removes the point source from the host galaxy without over- or under-subtraction. As a test, we apply this algorithm to images of artificial and $z \sim 0$--$1$ real AGNs observed with Hyper Suprime-Cam on the Subaru Telescope. We find that the spatial resolution of the host-galaxy images is improved to a level comparable to that of images from the Hubble Space Telescope and that the bright central point sources are removed. This algorithm is expected to enable statistical morphological studies of distant AGN host galaxies when applied to wide-field survey data from the Vera C. Rubin Observatory, the Euclid Space Telescope, and the Roman Space Telescope.
\end{abstract}


\section{Introduction}\label{sec:intro}

Galaxies and the supermassive black holes (SMBHs) at their centers are thought to be closely related in their growth and evolution \citep{2013ARA&A..51..511K, 2014ARA&A..52..589H}. Galaxy mergers are among the key physical mechanisms proposed to contribute to the coevolution of host galaxies and SMBHs over cosmic time. Theoretical and simulation studies predict that gas-rich major mergers can drive large amounts of gas toward galactic centers, thereby accelerating the SMBH growth \citep{2005Natur.433..604D, 2005ApJ...630..705H, 2006ApJS..163....1H}. So far, the connection between active galactic nucleus (AGN) activity and galaxy interactions or mergers at $z\gtrsim1$ has been investigated using high-resolution imaging observations with the Hubble Space Telescope (hereafter, Hubble) through morphological analyses of bright quasars with $M_{1450} \lesssim -24$ identified in ground-based surveys \citep[e.g.,][]{2015ApJ...806..218G, 2016ApJ...830..156M, 2019MNRAS.489..497Z, 2019ApJ...880..157D, 2023Natur.616...45C}. More recently, the James Webb Space Telescope (JWST) has begun to reveal the stellar and structural properties of quasar host galaxies at high redshifts \citep{2022ApJ...939L..28D, 2022ApJ...940L...7W, 2024A&A...689A.219D, 2025A&A...702A.174M, 2025ApJ...993...91D}, including the host galaxies of faint quasars with $M_{1450} \gtrsim -24$, discovered in the Hyper Suprime-Cam (HSC) wide-field surveys \citep{2018ApJ...869..150M, 2025NatAs...9.1541O, 2025ApJ...995...21T}. These morphological studies have reported mixed results on the connection between AGN activity and galaxy mergers. Some studies suggest that mergers play an important role in triggering luminous or obscured AGN activity \citep{2015ApJ...806..218G, 2018ApJ...853...63D, 2018PASJ...70S..37G, 2022ApJ...940L...7W, 2024A&A...689A.219D, 2025ApJ...978...74B, 2026ApJ...997...47B, 2026A&A...708A.373L}. In contrast, other studies have found no significant excess of disturbed structures in AGN host galaxies compared with inactive galaxies, suggesting that major mergers are not the dominant triggering mechanism for typical or even luminous AGN populations \citep{2011ApJ...726...57C, 2012ApJ...744..148K, 2017MNRAS.466..812V, 2021ApJ...919..129L, 2025ApJ...994..265V}. Taken together, the relationship between the AGN host-galaxy morphology and the AGN activity is not yet fully understood.

In morphological studies of the AGN host galaxy with Hubble and JWST, the central point sources are removed typically via two-dimensional image decomposition with GALFIT \citep{2002AJ....124..266P, 2010AJ....139.2097P}, in which the galaxy is modeled as an extended Sérsic component \citep{1963BAAA....6...41S} and the AGN as a point-source component. However, the narrow fields of view of Hubble and JWST limit statistical studies of the AGN host-galaxy morphology based on large samples of $ \gtrsim 10^{4}$ objects. In addition, the two-dimensional image decomposition is sensitive to uncertainties in the central region caused by point spread function (PSF) over-subtraction and to complex substructures in host galaxies \citep{2016ApJ...830..156M, 2025ApJ...993...91D}. Previous studies have also pointed out that uncertainty remains in the estimation of the S\'ersic index $n$ \citep{2008ApJ...683..644S}.

Another approach to studying the morphology of rare distant galaxies is to enhance the spatial resolution of images from ground-based telescopes using machine learning \citep[e.g.,][]{2017MNRAS.467L.110S, 2021arXiv210309711G, 2025PASJ...77...21S, 2026arXiv260420195K} or PSF deconvolution. In particular, PSF deconvolution is considered a powerful method because it does not require large training datasets used for machine learning. Traditionally, the Richardson--Lucy deconvolution has been widely used in astronomy \citep{1972JOSA...62...55R,1974AJ.....79..745L}. Since then, numerous PSF deconvolution algorithms have been proposed to accurately recover the morphologies of astronomical objects \citep[e.g.,][]{1998ApJ...494..472M, 2005ARA&A..43..139P, 2017A&A...601A..66F, 2019PASJ...71...24M, 2022PASJ...74.1329M, 2024AJ....167...96L}. Some deconvolution algorithms have separately processed a spatially extended component (e.g., diffuse gas) and a highly compact point-source component (e.g., pulsars) by imposing component-specific regularization terms \citep{2024PASJ...76..272M,2024AJ....168...55M}. However, these previous studies were not designed to provide the three reconstructed outputs required for our purpose: an extended component, a point-source component, and their combined reconstructed image. Obtaining the separated extended and point-source components as distinct output images is essential for morphological and photometric studies of faint AGN galaxies at high redshifts.

In this work, we develop a three-output PSF deconvolution method for AGN images obtained with Subaru/HSC. The method explicitly outputs an extended component corresponding to the host galaxy, a point-source component corresponding to the AGN, and a combined reconstructed image obtained by summing the two components. This approach simultaneously enhances the spatial resolution and removes the central point source, thereby enabling direct morphological analysis of the host-galaxy component. If the resolution enhancement and the point-source removal can be achieved for ground-based wide-field survey data through PSF deconvolution, it is possible to investigate the relationship between host-galaxy morphology and AGN activity using samples of $\gtrsim10^{4}$ objects. The proposed method would also reduce reliance on visual classification of AGN host galaxy morphologies, which has been applied in previous studies \citep{2015ApJS..221...11K, 2016ApJ...830..156M}. In addition, this method would enable morphological analyses based on non-parametric morphological indices, such as concentration, asymmetry, smoothness, the Gini coefficient, and $M_{20}$ \citep{2003ApJS..147....1C, 2004AJ....128..163L} by mitigating the AGN-contamination biases that have affected such measurements \citep{2022MNRAS.514..607G}.

This paper is organized as follows. In section \ref{sec:method}, we describe the deconvolution method and the optimization procedure. Section \ref{sec:data} presents the properties of the observational datasets. Section \ref{sec:objects} details AGNs used for the PSF deconvolution analysis. In Section \ref{sec:convergence}, we describe the convergence criterion and computational time for applying the deconvolution method. Section \ref{sec:result} presents the results and discusses the quality of reconstructed images through both qualitative and quantitative evaluations. Finally, section \ref{sec:conclusions} summarizes our findings and future prospects. All magnitudes are given in the AB system. We assume a flat $\Lambda$CDM cosmology with $H_0 = 70~\mathrm{km\,s^{-1}\,Mpc^{-1}}$, $\Omega_{\mathrm{m}} = 0.30$, and $\Omega_{\Lambda} = 0.70$.

\section{Deconvolution method}\label{sec:method}

\begin{figure}
 \begin{center}
  \includegraphics[width=8cm]{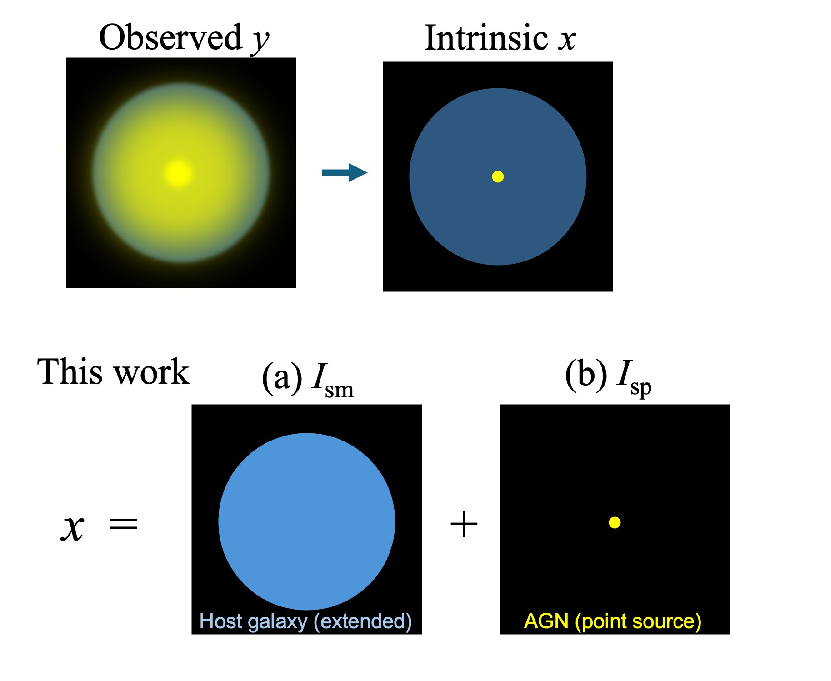} 
 \end{center}
\caption{Schematic overview of our method. Blue regions indicate a spatially extended component (e.g., an AGN host galaxy), while yellow regions in the images represent a point source (e.g., an AGN). (Upper panel) Conventional method: PSF deconvolution is applied to the blurred observed image $y$ (left) to obtain the intrinsic image $x$ (right). (Lower panel) Our method: The intrinsic image $x$ is decomposed into two components: (a) the extended component $\Ism$ and (b) the point-source component $\Isp$.}
\label{fig:overview}
\end{figure}

In this section, we describe our deconvolution method. Figure~\ref{fig:overview} shows a schematic overview of the PSF deconvolution. We analyze objects consisting of two components: an extended component and a point source. Conventional methods typically obtain the intrinsic image $x$ by applying PSF deconvolution to the blurred observed image $y$, as shown in the upper part of figure~\ref{fig:overview}. However, conventional methods are insufficient to analyze the detailed morphology of AGN host galaxies because AGN host galaxies at high-$z$ are compact and easily contaminated by flux from the central point source. To address this issue, our method separates the intrinsic image $x$ into two components: (1) a spatially extended component $\Ism$, i.e., a host-galaxy image (figure~\ref{fig:overview}a), and (2) a point-source component $\Isp$, i.e., an AGN image (figure~\ref{fig:overview}b).

In the following subsections, we describe the observational model in section~\ref{sec:obsmodel}, introduce the regularization terms in section~\ref{sec:regularization}, and present the optimization algorithm in section~\ref{sec:optimization}.

\subsection{Observation model}\label{sec:obsmodel}

We formulate the observation model for an astronomical object. We apply deconvolution to a small region of the celestial sphere. The objects in the target region are projected onto a rectangular region of $M = m \times n$ pixels on the tangent plane. Let $u = (i, j)$ ($i = 1, \dots, m; j = 1, \dots, n$) represent the coordinates on the celestial sphere, while $v$ ($v = 1, \dots, V$) denotes the pixel index on the detector. The observed flux value is given by $Y_v$. The PSF for the light emitted from the coordinates $u$ and detected at the pixel $v$ is characterized by the response matrix $t_{v, u}$. We define the intrinsic image $x$ as $x_u = \Ismpix{u} + \Isppix{u}$, where $\Ismpix{u}$ and $\Isppix{u}$ are the extended and point-source components at the pixel $u$, respectively (figure~\ref{fig:overview}). The pixel values $\Ismpix{u}$ and $\Isppix{u}$ are both non-negative real numbers.

In this observation model, we assume that the flux value $Y_v$ follows the Gaussian distribution shown below:

\begin{equation}
Y_v \sim \mathrm{Gauss}\!\left[
\sum_u t_{v,u}\left(\Ismpix{u} + \Isppix{u}\right),\, \sigma_v
\right],
\label{eq:obs_model}
\end{equation}

\noindent where $\sigma_v$ is the standard deviation of the noise at the pixel $v$. 

The likelihood is given by 

\begin{equation}
\begin{aligned}
P(Y \mid \Ism, \Isp) &= \prod_{v=1}^{V} \mathrm{Gauss}\!\left[ Y_v;\, \sum_{u=1}^{M} t_{v,u}(\Ismpix{u} + \Isppix{u}),\, \sigma_v \right] \\
&\hspace{-6.0em}= \prod_{v=1}^{V} \frac{1}{\sqrt{2\pi\sigma_v^2}} \exp\!\left\{ -\frac{\left[ Y_v - \sum_{u=1}^{M} t_{v,u}(\Ismpix{u} + \Isppix{u}) \right]^2}{2\sigma_v^2} \right\}.
\end{aligned}
\label{eq:likelihood}
\end{equation}

\subsection{Regularization terms}\label{sec:regularization}

To obtain the image $x$ closer to the intrinsic image, we apply regularization in the observation model. In this work, we introduce three regularization terms: (1) a smooth constraint on the image $\Ism$, which spatially smooths the host-galaxy structure; (2) a sparse constraint on the image $\Isp$, which localizes the point source to a small number of pixels; and (3) a new constraint, the point-source balance constraint, based on the pixel-wise product $\Ism \times \Isp$, which removes the point source from the host galaxy without over- or under-subtraction.

First, we apply the smooth constraint to the extended component $\Ismpix{u}$. Traditional deconvolution techniques use two-dimensional differences along the $i$- and $j$-directions to constrain adjacent pixels to have similar flux values \citep{Tikhonov:1963}. We define the prior distribution for the smooth constraint and the difference $V(\Ism)$ as follows:

\begin{equation}
p_{\mathrm{sm}}(\Ism) = Z_{\mathrm{sm}} \exp\!\left[-\alphasm V(\Ism)\right],
\label{eq:prior_smooth}
\end{equation}

\begin{equation}
\begin{aligned}
V(\Ism) &= \sum_{i=1}^{m-1} \sum_{j=1}^{n-1} \left[ \left(\Ismpix{i,j} - \Ismpix{i+1,j}\right)^2 + \left(\Ismpix{i,j} - \Ismpix{i,j+1}\right)^2 \right] \\
&\hspace{-3.0em}\quad + \sum_{i=1}^{m-1} \left(\Ismpix{i,n} - \Ismpix{i+1,n}\right)^2 + \sum_{j=1}^{n-1} \left(\Ismpix{m,j} - \Ismpix{m,j+1}\right)^2,
\end{aligned}
\label{eq:smoothness}
\end{equation}

\noindent where $Z_{\mathrm{sm}}$ is a normalization constant, and $\alphasm \geq 0$ is a hyperparameter representing the strength of the smooth constraint. As $\alphasm$ increases, the differences between adjacent pixels along the two-dimensional directions decrease, resulting in a smoother spatial structure. $V(\Ism)$ can be further expressed using the difference operators $V^{(x)}$ and $V^{(y)}$ as follows: 

\begin{equation}
V(\Ism)=\sum_w \left(\sum_u V^{(x)}_{w,u} \Ismpix{u}\right)^2 + \sum_w \left(\sum_u V^{(y)}_{w,u} \Ismpix{u}\right)^2.
\label{eq:smoothness_matrix}
\end{equation}

Next, we impose the sparse constraint on the point-source component $\Isppix{u}$. Traditional deconvolution techniques use sparse modeling with the $L_1$ norm, specifically the Least Absolute Shrinkage and Selection Operator (LASSO; \citealt{10.1111/j.2517-6161.1996.tb02080.x}), to allow only a small number of pixels to have non-zero flux values. We define the prior distribution for the sparse constraint and the $L_1$ norm $\|\Isp\|_1$ as follows:

\begin{equation}
p_{\mathrm{sp}}(\Isp)=Z_{\mathrm{sp}}\exp\!\left[-\alphasp \|\Isp\|_1\right],
\label{eq:prior_sparse}
\end{equation}

\begin{equation}
\|\Isp\|_1 = \sum_u |\Isppix{u}|,
\label{eq:l1norm}
\end{equation}

\noindent where $Z_{\mathrm{sp}}$ is a normalization constant, and $\alphasp \geq 0$ is a hyperparameter representing the strength of the sparse constraint. As $\alphasp$ increases, the effect of the sparse constraint becomes stronger, reducing the number of pixels with non-zero flux values.

\begin{figure*}
 \begin{center}
  \includegraphics[width=16cm]{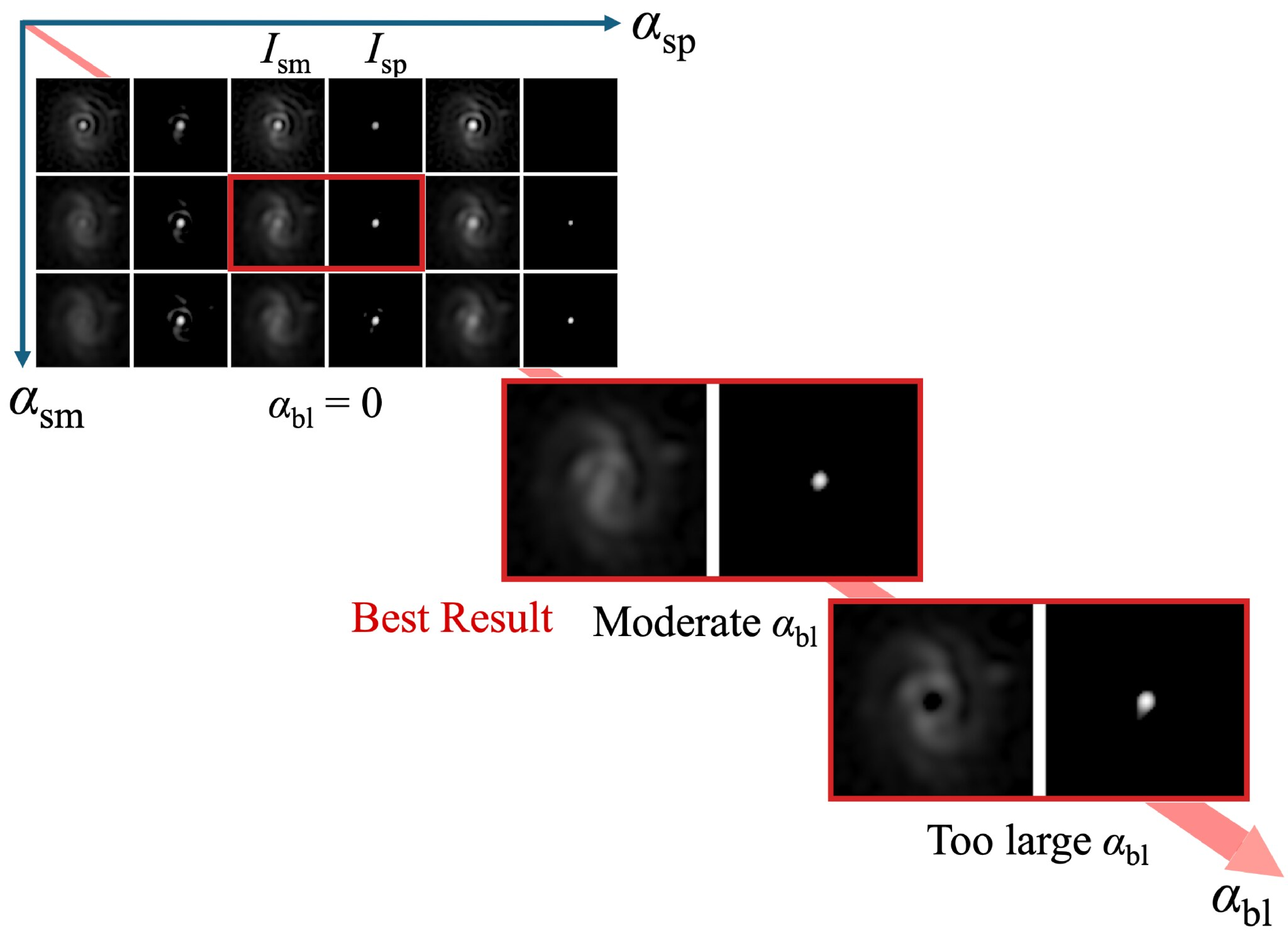}
 \end{center}
\caption{Dependence of the reconstructed images for an AGN host galaxy on the hyperparameters. The upper-left panels show the reconstructed images in the $\alphasm$ and $\alphasp$ space. The red diagonal axis corresponds to the hyperparameter $\alphabl$ for the point-source balance constraint. See section \ref{sec:regularization} for details. In each set, the left and right images indicate the extended and point-source components, respectively.}
\label{fig:hyper3d}
\end{figure*}

Figure~\ref{fig:hyper3d} shows the dependence of the reconstructed images on the hyperparameters $\alphasm$ and $\alphasp$ for an AGN. The details of this object are described in section~\ref{sec:objects}. As $\alphasm$ increases and the smooth constraint becomes stronger, the extended component $\Ism$ (i.e., the host galaxy) becomes smoother. Similarly, as $\alphasp$ increases and the sparse constraint becomes stronger, fewer pixels in the point-source component $\Isp$ (i.e., the AGN) have non-zero flux values. However, the reconstruction of one component also affects the result of the other. This is because the reconstructed image $x$ is expressed as the sum of $\Ism$ and $\Isp$ ($x = \Ism + \Isp$; see section~\ref{sec:obsmodel}). For example, when $\alphasp$ is small, as $\alphasm$ increases, fine-scale components that can no longer be captured by the smooth-extended component $\Ism$ ``leak'' into the point-source component $\Isp$. Due to this interdependence of the hyperparameters $\alphasm$ and $\alphasp$ for the $\Ism$ and $\Isp$ components, it is difficult to find an appropriate combination of $\alphasm$ and $\alphasp$, even after careful adjustment. In fact, even with the best combination of $\alphasm$ and $\alphasp$ in the $\alphasm$--$\alphasp$ space, point-source flux still remains in the host-galaxy component (indicated by the red box in the upper-left corner of figure~\ref{fig:hyper3d}).

To better separate the extended structure and the point source, we introduce the point-source balance constraint. We define the prior distribution for the point-source balance constraint and the product $W(\Ism, \Isp)$ of the extended component $\Ism$ and the point-source component $\Isp$ as follows:

\begin{equation}
p_{\mathrm{bl}}(\Ism,\Isp) = Z_{\mathrm{bl}}\exp\!\left[-\alphabl W(\Ism,\Isp)\right],
\label{eq:prior_balance}
\end{equation}

\begin{equation}
W(\Ism,\Isp)=\sum_u \Ismpix{u} \Isppix{u},
\label{eq:balance_term}
\end{equation}

\noindent where $Z_{\mathrm{bl}}$ is the normalization constant and $\alphabl \ge 0$ is a hyperparameter representing the balance between the flux values of the extended component $\Ism$ and the point-source component $\Isp$. The product $W(\Ism, \Isp)$ takes the form of a cross-penalty based on the product of these two components, $\Ism \times \Isp$. As $\alphabl$ increases, this constraint penalizes the simultaneous presence of large flux values in the point-source and extended components at the same pixel location in each image. For example, at pixels where the point-source component $\Isp$ has a large flux value, increasing $\alphabl$ reduces the flux value of the extended component $\Ism$ at the corresponding pixel location. The red axis in figure~\ref{fig:hyper3d} shows the dependence of the $\Isp$ and $\Ism$ images on $\alphabl$. A moderate $\alphabl$ value produces a reconstructed $\Ism$ image without over- or under-subtraction of the point source. Note that the central flux of the extended component can be excessively suppressed when $\alpha_{\mathrm{bl}}$ is too large (the lower right panel in figure~\ref{fig:hyper3d}).

\subsection{Optimization}\label{sec:optimization}

We present the objective function derived from the maximization of the posterior distribution and the update equations. Using the observation model and the three regularization terms introduced in the previous sections, we simultaneously estimate the extended component $\Ism$ and the point-source component $\Isp$. First, when the model used in this work is interpreted probabilistically, the model can be formulated as the maximization of the posterior probability distribution. Specifically, the posterior distribution consists of the likelihood of the observation model and the prior distributions corresponding to each regularization term. The posterior distribution $P(\Ism, \Isp \mid Y)$ is given by

\begin{equation}
\begin{aligned}
P(\Ism,\Isp \mid Y)&=\frac{p(\Ism,\Isp,Y)}{p(Y)}\\
&\hspace{-4.5em}=\frac{p_{\mathrm{sm}}(\Ism)\,p_{\mathrm{sp}}(\Isp)\,p_{\mathrm{bl}}(\Ism,\Isp)\,p(Y \mid \Ism,\Isp)}{p(Y)}.
\end{aligned}
\label{eq:posterior}
\end{equation}

\noindent where $p$ denotes a probability distribution. The subscripts $\mathrm{sm}$, $\mathrm{sp}$, and $\mathrm{bl}$ indicate the prior distributions corresponding to the smooth, sparse, and point-source balance constraints, respectively.

\noindent The logarithm of the posterior distribution is expressed as the sum of terms as follows:

\begin{equation}
\begin{split}
\log p(\Ism,\Isp \mid Y)
&= \log p(Y \mid \Ism,\Isp)+\log p_{\mathrm{sm}}(\Ism) \\[3pt]
&\hspace{-4.5em}\quad +\log p_{\mathrm{sp}}(\Isp)+\log p_{\mathrm{bl}}(\Ism,\Isp)+\mathrm{const.} \\[4pt]
&\hspace{-6.5em}= -\sum_v \frac{1}{2\sigma_v^2}\left[ Y_v-\sum_u t_{v,u}\left(\Ismpix{u}+\Isppix{u}\right) \right]^2 \\[4pt]
&\hspace{-6.5em}\quad -\alphasm V(\Ism)-\alphasp \|\Isp\|_1-\alphabl W(\Ism,\Isp)+\mathrm{const.}
\end{split}
\label{eq:logposterior}
\end{equation}

In our algorithm, the image reconstruction is performed by minimizing the negative log-posterior distribution. We formulate the following optimization problem to minimize the objective function:

\begin{align}
\mathop{\mathrm{arg\,min}}\limits_{\Ismpix{u},\,\Isppix{u} \ge 0} \Biggl\{
&\sum_v \frac{1}{2\sigma_v^2}\left[\sum_u t_{v,u}\left(\Ismpix{u}+\Isppix{u}\right)-Y_v\right]^2 \notag\\[4pt]
&\hspace{-5.5em}\quad + \alphasm\sum_w \left[\sum_u V^{(x)}_{w,u}\Ismpix{u}\right]^2
      + \alphasm\sum_w \left[\sum_u V^{(y)}_{w,u}\Ismpix{u}\right]^2 \notag\\[4pt]
&\quad + \alphasp \|\Isp\|_1 + \alphabl\sum_u \Ismpix{u}\Isppix{u}
\Biggr\}.
\label{eq:map_estimator}
\end{align}

\noindent The objective function is decomposed into a continuously differentiable term $f(\Ism, \Isp)$ and a non-differentiable regularization term $g(\Isp)$ as follows:

\begin{align}
f(\Ism,\Isp)
&= \sum_v \frac{1}{2\sigma_v^2}\left[\sum_u t_{v,u}\left(\Ismpix{u}+\Isppix{u}\right)-Y_v\right]^2 \notag\\[4pt]
&\hspace{-5.5em}\quad + \alphasm\sum_w \left[\sum_u V^{(x)}_{w,u}\Ismpix{u}\right]^2
      + \alphasm\sum_w \left[\sum_u V^{(y)}_{w,u}\Ismpix{u}\right]^2 \notag\\[4pt]
&\quad + \alphabl\sum_u \Ismpix{u}\Isppix{u},
\label{eq:objective_smooth}
\end{align}

\begin{equation}
g(\Isp)=\alphasp \|\Isp\|_1.
\label{eq:g_sparse}
\end{equation}

\noindent This minimization problem containing the non-differentiable term $g(\Isp)$ cannot be solved by standard gradient descent. We solve this problem using the Iterative Shrinkage-Thresholding Algorithm (ISTA; \citealp{article, articles}), which is an effective method for expressions of the form $f(\Ism, \Isp) + g(\Isp)$. In the ISTA, the differentiable term $f(\Ism, \Isp)$ is approximated to first order around the current iteration point. We then introduce an approximation function obtained by adding a quadratic stabilization term with the Lipschitz constant $L_k$. The approximation function $R(\Ism, \Isp)$ is given by

\begin{align}
R(\Ism,\Isp)
&= g(\Isp)+\frac{L_k}{2}\Bigl\|(\Ism,\Isp)^T-(\Ism^{(k-1)},\Isp^{(k-1)})^T \notag\\
&\qquad\qquad + \frac{1}{L_k}\nabla_{\Ism,\Isp}f\!\left(\Ism^{(k-1)},\Isp^{(k-1)}\right)\Bigr\|^2.
\label{eq:prox_objective}
\end{align}

At each iteration of the ISTA, the extended component $\Ismpix{u}$ and the point-source component $\Isppix{u}$ are estimated simultaneously using the approximation function $R(\Ism, \Isp)$ in equation~(\ref{eq:prox_objective}). In this process, the update equations for the $\Ism$ and $\Isp$ components are derived independently. Here we define the variables $a_u^{(k-1)}$ and $b_u^{(k-1)}$ as follows:

\begin{equation}
a_u^{(k-1)}=\Ismpix{u}^{(k-1)}-\frac{1}{L_k}\left[\nabla_{\Ism} f\!\left(\Ism^{(k-1)},\Isp^{(k-1)}\right)\right]_u,
\label{eq:def_a}
\end{equation}

\begin{equation}
b_u^{(k-1)}=\Isppix{u}^{(k-1)}-\frac{1}{L_k}\left[\nabla_{\Isp} f\!\left(\Ism^{(k-1)},\Isp^{(k-1)}\right)\right]_u.
\label{eq:def_b}
\end{equation}

Using these variables $a$ and $b$, we derive the following update equations.

\begin{equation}
\Ismpix{u}^{(k)}=\max\!\left(a_u^{(k-1)},\,0\right),
\label{eq:update_I}
\end{equation}

\begin{equation}
\Isppix{u}^{(k)}=
\begin{cases}
\max\!\left(b_u^{(k-1)}-\dfrac{\alphasp}{L_k},\,0\right) & b_u^{(k-1)} \ge 0,\\[4pt]
0 & b_u^{(k-1)} < 0.
\end{cases}
\label{eq:update_S}
\end{equation}

\noindent To ensure stable solution updates, we introduce the backtracking scheme for the Lipschitz constant $L_k$ in the ISTA. The gradient $\nabla f$ of the differentiable term $f(\Ism, \Isp)$ used in our algorithm is assumed to be Lipschitz continuous. Under this assumption, the quadratic approximation function $Q(\Ism, \Isp)$, defined with the Lipschitz constant $L$, serves as an upper bound for $f(\Ism, \Isp)$. The condition for this upper bound property to hold is given as follows:

\begin{equation}
f\!\left(\Ism^{(k)},\Isp^{(k)}\right) \le Q_{L_k}^{(f)}\!\left[(\Ism^{(k)},\Isp^{(k)}),(\Ism^{(k-1)},\Isp^{(k-1)})\right].
\label{eq:backtracking_condition}
\end{equation}

At each iteration, the Lipschitz constant $L_k$ satisfying the condition is determined by a backtracking scheme. Specifically, $L_k$ is initialized to $L_{k-1}$ and then increased until the upper-bound condition is satisfied. Through this process, the objective function value is guaranteed to decrease monotonically at each iteration of the ISTA. This monotonic decrease enables stable optimization, allowing for the subsequent recovery of the host-galaxy structure and the effective removal of the point source.

\section{Data}\label{sec:data}

\begin{longtable}{cccccc}
  \caption{Galaxies with AGN used for the analysis}\label{tab:target_galaxies}  
\hline\noalign{\vskip3pt} 
  Object name & Nickname & RA (deg) & Dec (deg) & $z_{\mathrm{phot}}$ & $i_{\mathrm{mag}}$ \\
  (1) & (2) & (3) & (4) & (5) & (6) \\   [2pt] 
\hline\noalign{\vskip3pt} 
\endfirsthead      
\hline\noalign{\vskip3pt} 
  Object name & Nickname & RA (deg) & Dec (deg) & $z_{\mathrm{phot}}$ & $i_{\mathrm{mag}}$ \\  [2pt] 
\hline\noalign{\vskip3pt} 
\endhead
\hline\noalign{\vskip3pt} 
\endfoot
\hline\noalign{\vskip3pt} 
\multicolumn{6}{@{}l@{}}{\hbox to0pt{\parbox{160mm}{\footnotesize
\hangindent6pt\noindent
Note. --- Columns: (1) Object name. (2) Object nickname. (3) Right Ascension (J2000.0) in degrees. \\ (4) Declination (J2000.0) in degrees. (5) Photometric redshift. (6) HSC $i$-band total magnitude.
}\hss}} 
\endlastfoot 
  COSMOS J095928.32+022106.9 & J0959 & $149.868020$ & $+2.351862$ & $0.34$ & $18.74$ \\
  COSMOS J100009.02+014112.4 & J1000 & $150.037569$ & $+1.686737$ & $1.38$ & $22.06$ \\
\end{longtable}

We describe details of objects and observational datasets for the deconvolution analysis. Using our deconvolution method, we analyze AGNs in $i$-band images from HSC \citep{2018PASJ...70S...1M}, a $1.8~\mathrm{deg}^2$ field-of-view optical camera on the Subaru Telescope. The goal is to enhance the spatial resolution of the AGN host-galaxy images to a level comparable to that of F814W-band images taken with the Advanced Camera for Surveys (ACS) on Hubble and remove the central point-source component. 

For the observed images, we use data from the HSC-Subaru Strategic Program (SSP) Data Release 3 (DR3; \citealt{2022PASJ...74..247A}). The $i$-band limiting magnitude of the HSC-SSP DR3 data is approximately 27.4~mag ($5\sigma$, $2^{\prime\prime}$ aperture) in the UltraDeep layer. The pixel scale of the HSC data is $0^{\prime\prime}\!.168~\mathrm{pixel}^{-1}$, and the PSF FWHM is typically between $0^{\prime\prime}\!.6$ and $1^{\prime\prime}\!.0$. For validation, we use Hubble ACS images from the COSMOS-Wide field \citep{2007ApJS..172..196K}. The limiting magnitude of these Hubble images is approximately 27.2~mag ($5\sigma$, $0.24^{\prime\prime}$ aperture). The pixel scale and the PSF FWHM of the Hubble data are $0^{\prime\prime}\!.03~\mathrm{pixel}^{-1}$ and $\sim 0^{\prime\prime}\!.1$, respectively.

The inputs to the PSF deconvolution analysis are the HSC images ($Y_v$), the corresponding PSF images used to construct the response matrix ($t_{v,u}$) in equation~(\ref{eq:obs_model}), and the variance images ($\sigma_v^2$) in equation~(\ref{eq:likelihood}). The PSF images are obtained from the COSMOS UltraDeep field using the HSC-SSP DR3 PSF picker. \footnote{\texttt{https://hsc-release.mtk.nao.ac.jp/psf/pdr3/}} The original PSF images have sizes of approximately $40 \times 40$ pixels and are zero-padded to match the size of each input image before deconvolution. The Hubble images are used as reference images to assess the reconstruction performance.

\section{AGNs for the analysis}\label{sec:objects}

In this work, we analyze two types of AGN data: artificial and real AGNs. The artificial AGN is constructed to mimic HSC observations. In section~\ref{sec:artificial}, we describe how the artificial AGN images are generated. In section~\ref{sec:real}, we detail the data acquisition procedure and properties of the real AGN images.

\subsection{Artificial AGN}\label{sec:artificial}

The artificial AGN image is generated as follows. First, a host galaxy is constructed using the S\'ersic profile \citep{1963BAAA....6...41S}:

\begin{equation}
I_{\mathrm{host}}(r)
= I_{\rm e} \exp\!\left\{-b_n\left[\left(\frac{r}{r_{\rm e}}\right)^{1/n}-1\right]\right\},
\label{eq:sersic}
\end{equation}

\noindent where $I_{\mathrm{host}}(r)$ represents the flux distribution of the host galaxy, $r_{\rm e}$ is the effective radius enclosing half of the total flux, $I_{\rm e}$ is the flux at $r_{\rm e}$, and $n$ is the S\'ersic index. We adopt a disk galaxy with a S\'ersic index of $n=1$ and an effective radius of $r_{\rm e}=12$~pixels, which corresponds to $\sim2$" in the HSC images. Next, a bright point source is added at the center of the generated host galaxy. The flux of the point-source component is set to 12\% of that of the host galaxy. These values are chosen so that the apparent size and the point-source-to-host flux ratio of the artificial AGN are roughly comparable to those of the real AGNs analyzed in this work.

An artificial AGN image of $150 \times 150$~pixels is then generated from this artificial galaxy model. The artificial galaxy is convolved with the HSC PSF to match the resolution of the HSC observed images, and Gaussian noise is added over the entire image. The noise level is adjusted so that the resulting image has the same signal-to-noise ratio ($S/N$) as the real AGN image described in section~\ref{sec:real} (specifically $S/N=398$). For the variance image, we adopt a uniform variance across the entire image, with the value determined by referring to the real AGN data to ensure a comparable noise level. The same PSF used for the convolution is also used for the PSF deconvolution analysis.

\subsection{Real AGN}\label{sec:real}

\begin{figure}
 \begin{center}
  \includegraphics[width=8cm]{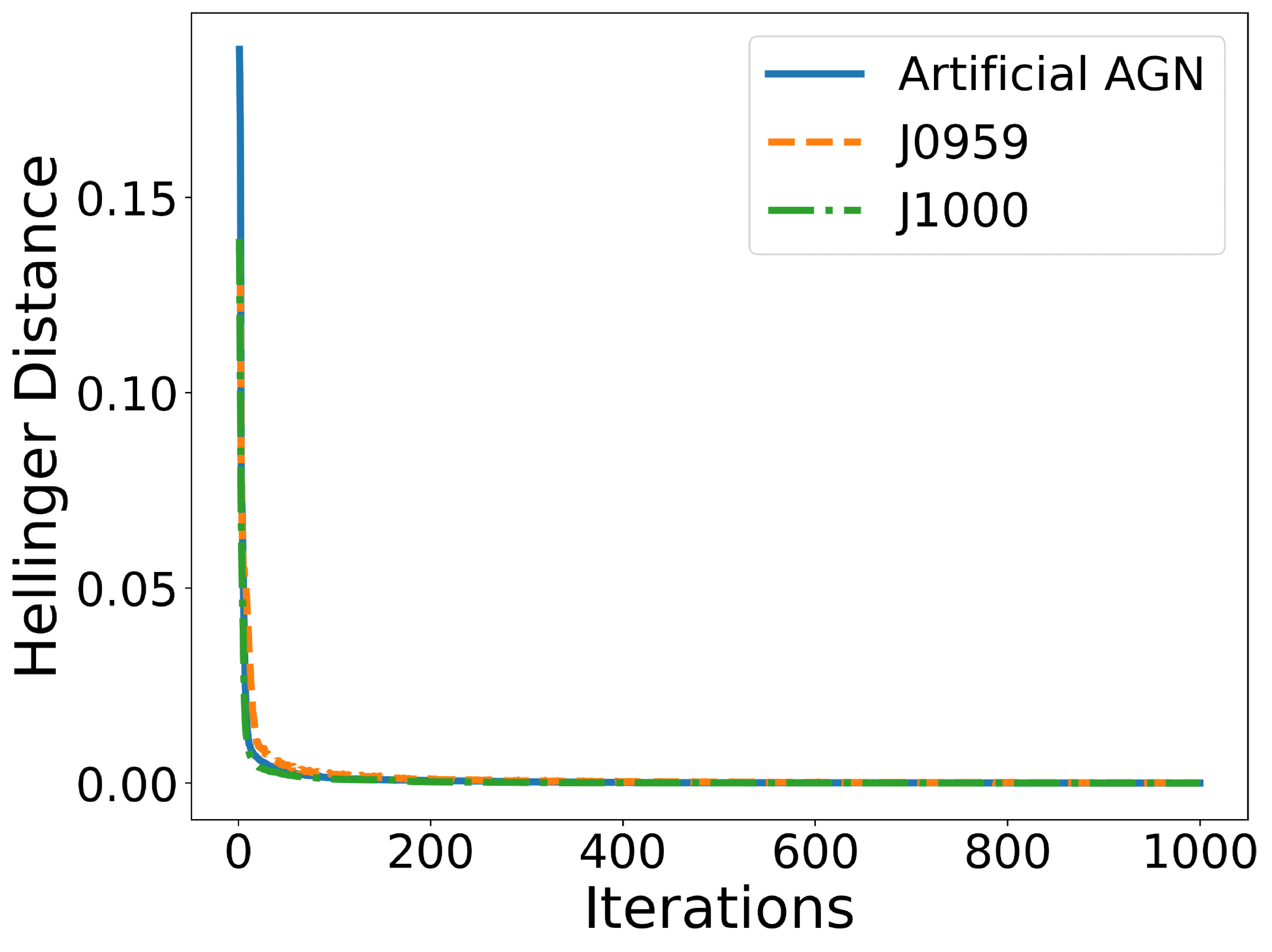} 
 \end{center}
\caption{Hellinger distance at each iteration for the three targets: the artificial AGN (the blue line), J0959 (the orange dashed line), and J1000 (the green dot-dashed line). See section~\ref{sec:convergence} for details regarding the definition and calculation of the Hellinger distance.}
\label{fig:hellinger}
\end{figure}

We describe the data acquisition procedure and the properties of the real AGNs. We analyze two real AGNs in the COSMOS field: COSMOS J095928.32+022106.9 (hereafter J0959) at $z=0.34$ and COSMOS J100009.02+014112.4 (hereafter J1000) at $z=1.38$. J0959 was previously studied as an AGN \citep{2009ApJ...691..705G}, Whereas J1000 was selected as a Chandra-detected X-ray source from the COSMOS2020 catalog \citep{2022ApJS..258...11W}. These objects are selected because both a central point source and characteristic host-galaxy features, such as spiral arms, are clearly shown, making them suitable objects for evaluating the performance of our deconvolution method. Basic information on the objects, including the coordinates and magnitudes, is summarized in table~\ref{tab:target_galaxies}.

The HSC observed, variance, and PSF images of the real AGNs are obtained from the DR3 database. The observed and variance images are cut out to include the target object and its surrounding host-galaxy structures. The cutout sizes are set to $15^{\prime\prime}$ for J0959 and $10^{\prime\prime}$ for J1000, so as to fully encompass the visible extent of the host galaxy. To ensure that the hyperparameter $\alphasp$ does not require substantial adjustment across different types of objects, the cutout images are resized to $150 \times 150$ pixels, matching the size of the artificial AGN images.\footnote{Because interpolating the flux values may alter the noise distribution and affect the image reconstruction results, we also analyze the original images without the pixel interpolation. This supplementary analysis finds no significant differences between the reconstruction results obtained with and without the pixel interpolation (see Appendix 1).} During this process, the pixel values are resampled onto a $150 \times 150$ grid using linear interpolation. For the PSF images, we use those obtained from the same field as the observed objects.

The Hubble images are obtained from the IRSA COSMOS cutouts.\footnote{\texttt{https://irsa.ipac.caltech.edu/data/COSMOS/index\_cutouts.html}} The retrieved images are resized to match the HSC pixel scale.

\section{Application to the AGN images}\label{sec:convergence}

We apply our deconvolution method to the three AGN images described in section~\ref{sec:objects}. Following equations~(\ref{eq:update_I}) and (\ref{eq:update_S}), the extended component $\Ism$ and the point-source component $\Isp$ are updated iteratively. The Hellinger distance is utilized as the convergence criterion for these image updates:

\begin{equation}
H = \frac{1}{\sqrt{2}} \sqrt{\sum_{i} \left( \sqrt{p_i} - \sqrt{q_i} \right)^2},
\label{eq:hellinger}
\end{equation}

\begin{figure*}
 \begin{center}
  \includegraphics[width=16cm]{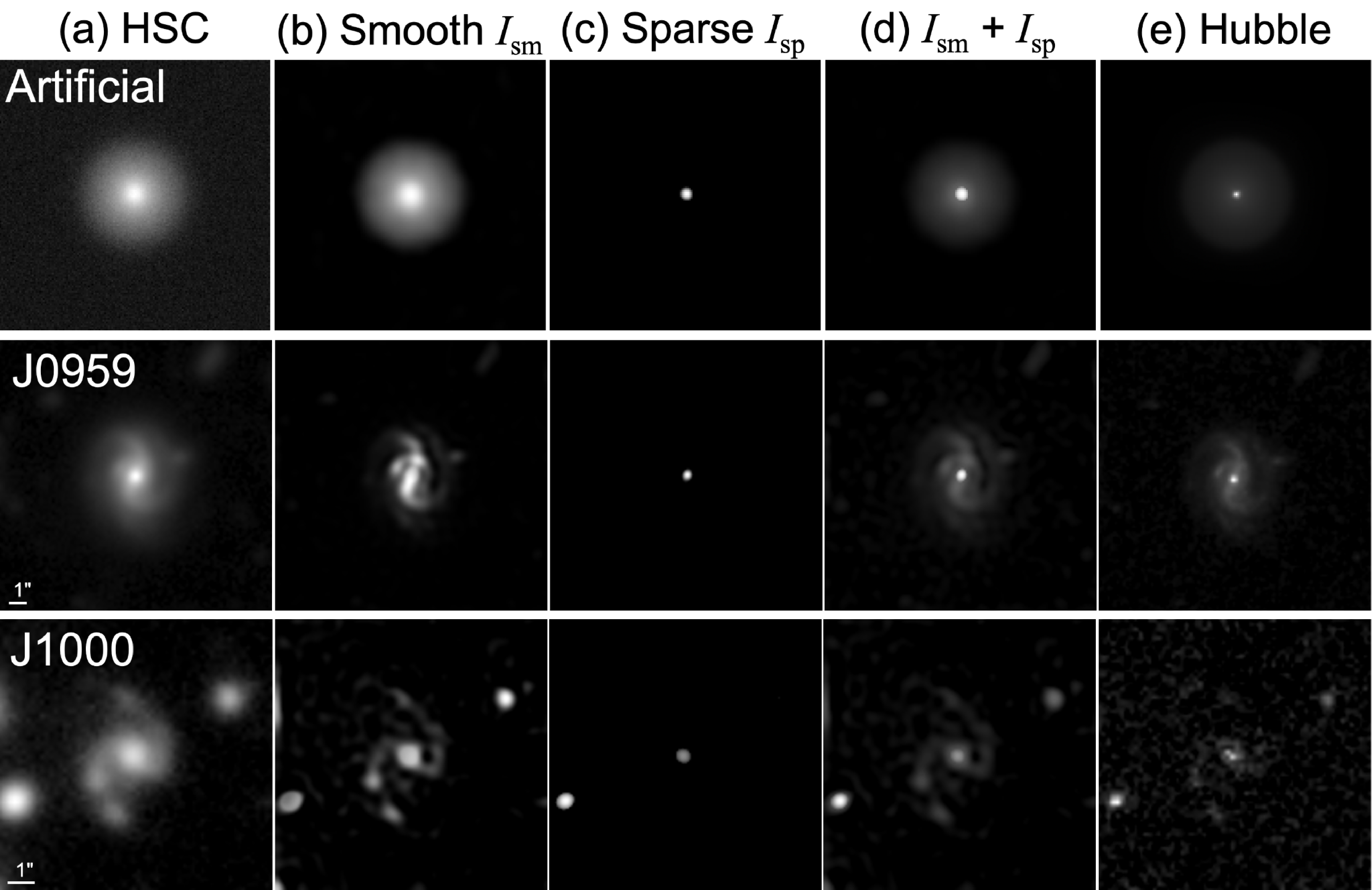} \\
 \end{center}
\caption{Comparison of the original HSC images, the reconstructed images, and the ground-truth Hubble images. The top row shows the artificial AGN, while the middle and bottom rows show the real AGNs, J0959 ($z = 0.34$) and J1000 ($z = 1.38$), respectively. In each row from left to right, the panels display: (a) the HSC image, (b) the smooth component $\Ism$, (c) the sparse component $\Isp$, (d) the combined reconstructed image $\Ism + \Isp$, and (e) the Hubble image. The model image of the artificial AGN is convolved with the Hubble PSF to match the spatial resolution of the ground-truth real AGN images. The white solid line in the leftmost panels of the real AGNs indicates a scale of $1^{\prime\prime}$.}
\label{fig:images_agn}
\end{figure*}

\noindent where $p_i$ and $q_i$ are the normalized probability distributions of the pixel values before and after the updates, respectively. The Hellinger distance is a metric used to quantify the difference between the reconstructed images obtained before and after each update. A smaller Hellinger distance indicates a smaller difference between the images. Figure~\ref{fig:hellinger} shows the Hellinger distance at each iteration. For all the analyzed images, the Hellinger distance decreases rapidly during the early iterations and then more gradually at later stages. After approximately 100 iterations, the change becomes small, suggesting that the iterative solution is sufficiently stable at this stage. To ensure convergence, all the images are processed for 1000 iterations. The computation time for 1000 iterations is approximately 7--8~seconds on an Apple M4 CPU.

We explore the best sets of hyperparameters for the smooth constraint $\alphasm$, the sparse constraint $\alphasp$, and the point-source balance constraint $\alphabl$ for each object via a grid search and visually compare the resulting output images. All output images are provided in Appendix 2. The best sets of hyperparameters $(\log_{10}\alphasm, \log_{10}\alphasp, \log_{10}\alphabl)$ are $(11.30, 7.11, 9.00)$ for the artificial AGN, $(12.60, 9.68, 12.08)$ for J0959, and $(10.00, 7.32, 10.30)$ for J1000. The dependence of the hyperparameters on object properties, such as magnitude and size, will be investigated in future work. 

We calculate the root-mean-square error (RMSE) for the final output image:

\begin{equation}
RMSE = \sqrt{\frac{1}{N} \sum_{i=1}^{N} (y_i - \hat{y}_i)^2},
\label{eq:RMSE}
\end{equation}

\noindent where $y_i$ is the pixel value of the reference image, $\hat{y}_i$ is the corresponding pixel value of the compared image, and $N$ is the total number of pixels. In this work, we use two types of RMSE. The first metric is $RMSE_{\mathrm{HSC-Conv}(\Ism+\Isp)}$, which is defined by comparing the HSC observed image with the reconstructed image $(\Ism+\Isp)$ convolved with the HSC PSF. This metric represents the fidelity of the reconstruction to the HSC observation. It indicates how well the reconstructed components reproduce the observed HSC image after convolution. The second metric is $RMSE_{\mathrm{Hubble}-(\Ism+\Isp)}$, which is defined by comparing the Hubble image with the reconstructed image $(\Ism+\Isp)$. This metric represents the reconstruction accuracy with respect to the ground-truth Hubble image. The calculated $RMSE$ values are summarized in table~\ref{tab:rmse}. The values of $RMSE_{\mathrm{HSC-Conv}(\Ism+\Isp)}$ are as small as $\sim 4\times10^{-6}$ to $2\times10^{-5}$, indicating that the reconstructed images reproduce the HSC observed images with high accuracy. The values of $RMSE_{\mathrm{Hubble}-(\Ism+\Isp)}$ are also relatively small, at $\sim 2\times10^{-2}$ to $3\times10^{-2}$, although these values are larger than the values of $RMSE_{\mathrm{HSC-Conv}(\Ism+\Isp)}$. This discrepancy is likely attributable to factors such as the difference in imaging depth between the HSC and Hubble data.

\begin{table}
  \tbl{RMSE values of reconstructed images\label{tab:rmse}}{%
  \begin{tabular}{lcc}
    \hline
    Object & $\mathrm{RMSE}_{\mathrm{HSC}\text{-}\mathrm{Conv}(\Ism+\Isp)}$ & $\mathrm{RMSE}_{\mathrm{Hubble}\text{-}(\Ism+\Isp)}$ \\
    (1) & (2) & (3) \\
    \hline
    Artificial AGN & $1.63 \times 10^{-5}$ & $3.34 \times 10^{-2}$ \\
    J0959 & $3.88 \times 10^{-6}$ & $1.67 \times 10^{-2}$ \\
    J1000 & $5.85 \times 10^{-6}$ & $2.17 \times 10^{-2}$ \\
    \hline
  \end{tabular}}
  \begin{tabnote}
    Note. --- Columns: (1) Object nickname. (2) RMSE between the reconstructed image convolved with the PSF and the HSC observed image. (3) RMSE between the reconstructed image and the Hubble image.
  \end{tabnote}
\end{table}

\begin{figure}
 \begin{center}
  \includegraphics[width=8cm]{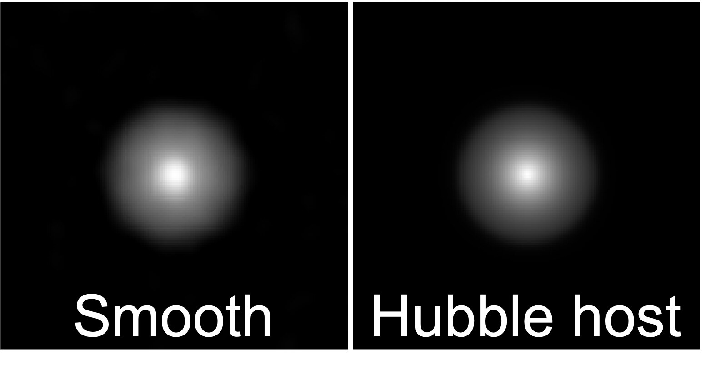} \\
 \end{center}
\caption{Comparison of the reconstructed host galaxy (left) and the Hubble PSF-convolved host-galaxy image (right) for the artificial AGN.}
\label{fig:images_host}
\end{figure}

\section{Results \& Discussion}\label{sec:result}

\begin{figure*}
 \begin{center}
  \includegraphics[width=7cm]{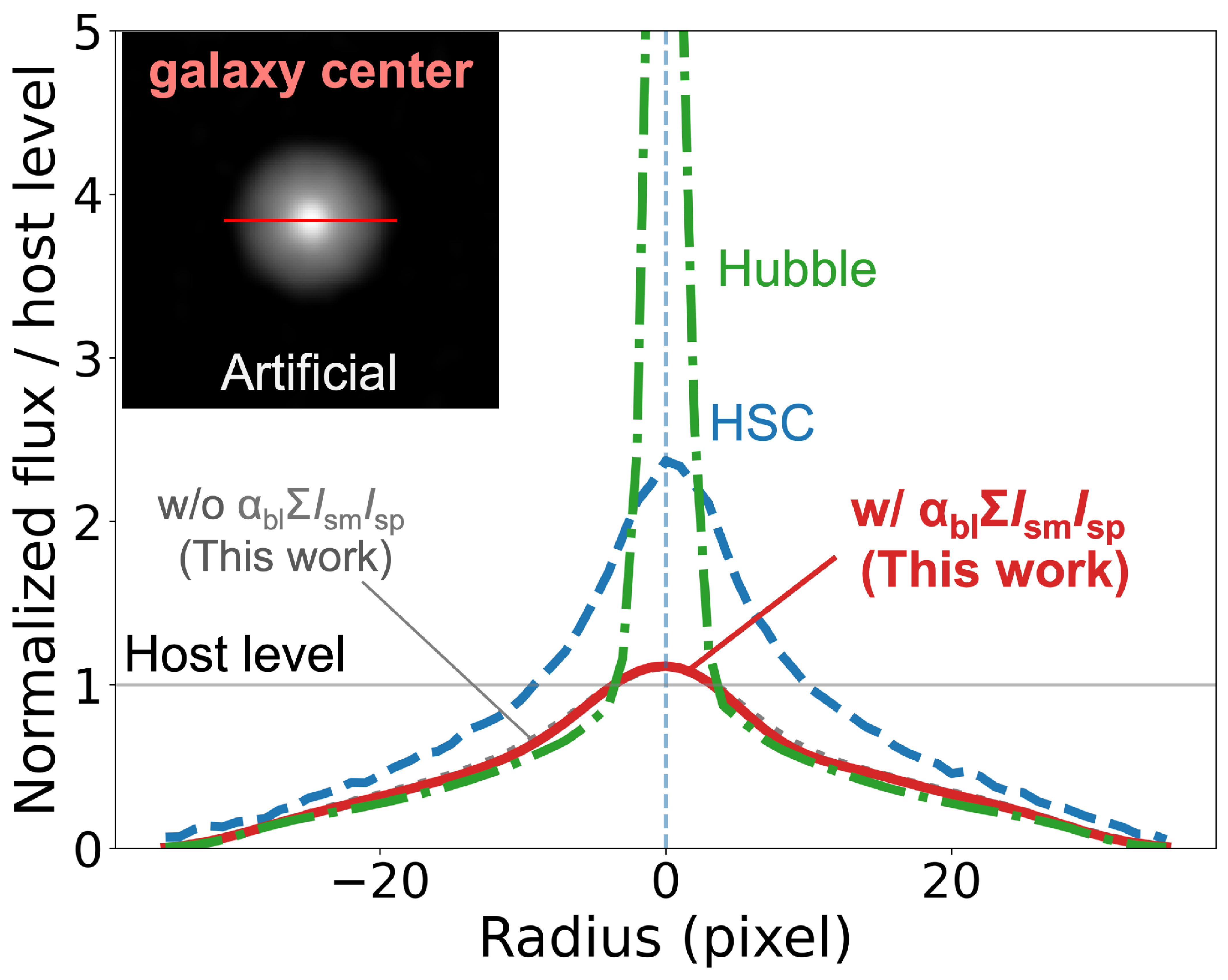} 
  \makebox[7cm][c]{\rule{0pt}{5cm}} \\
  \includegraphics[width=7cm]{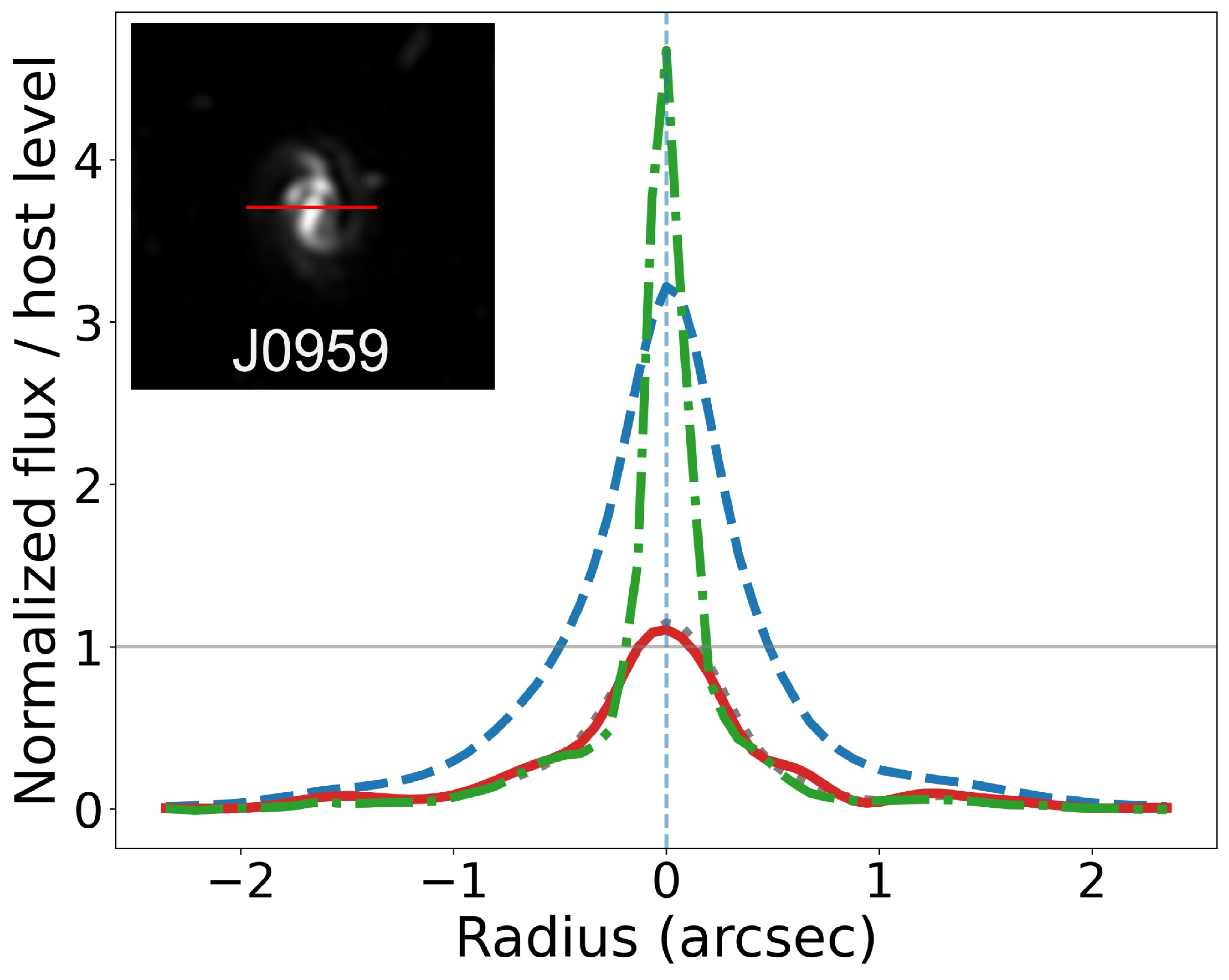} 
  \includegraphics[width=7cm]{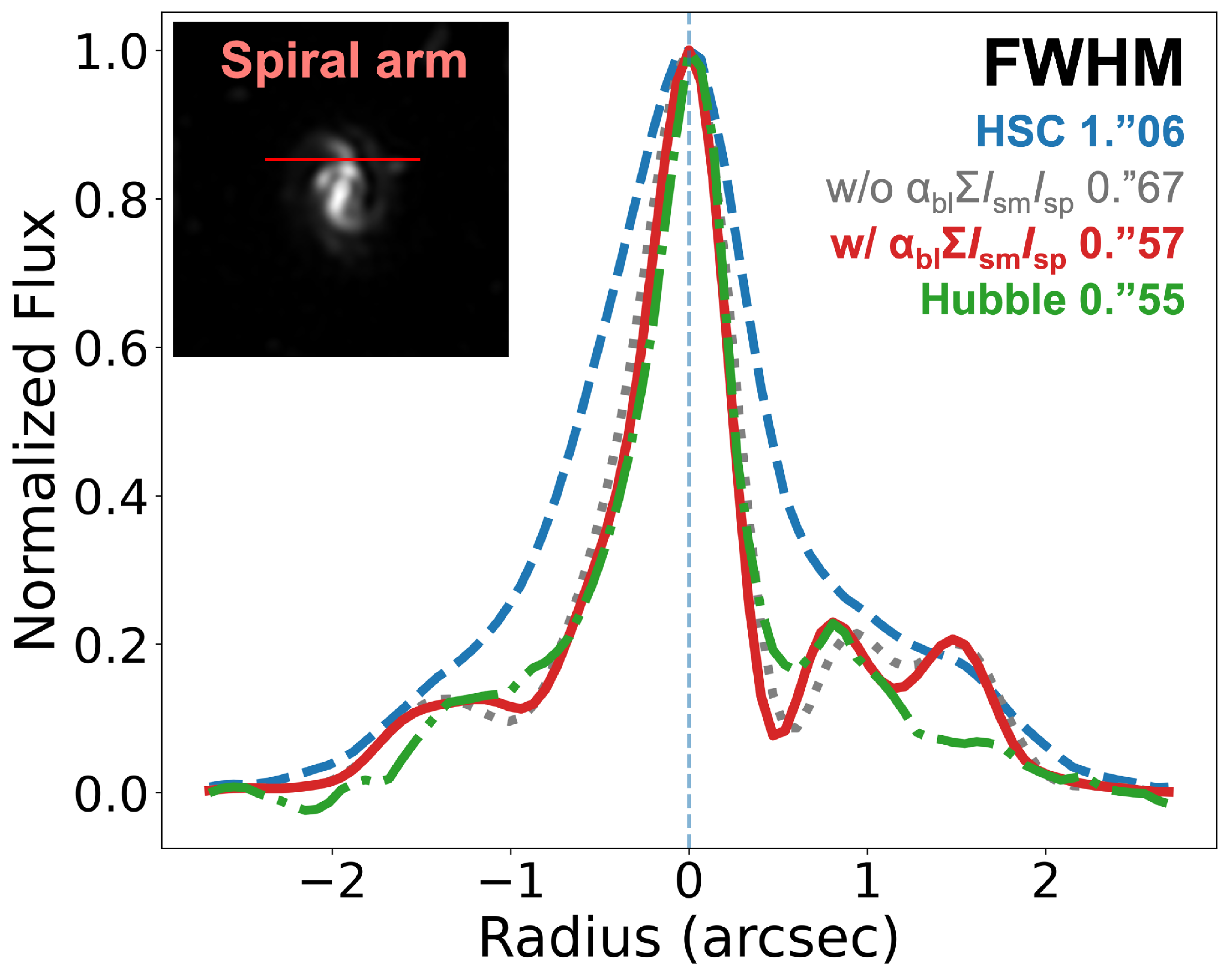} \\
  \includegraphics[width=7cm]{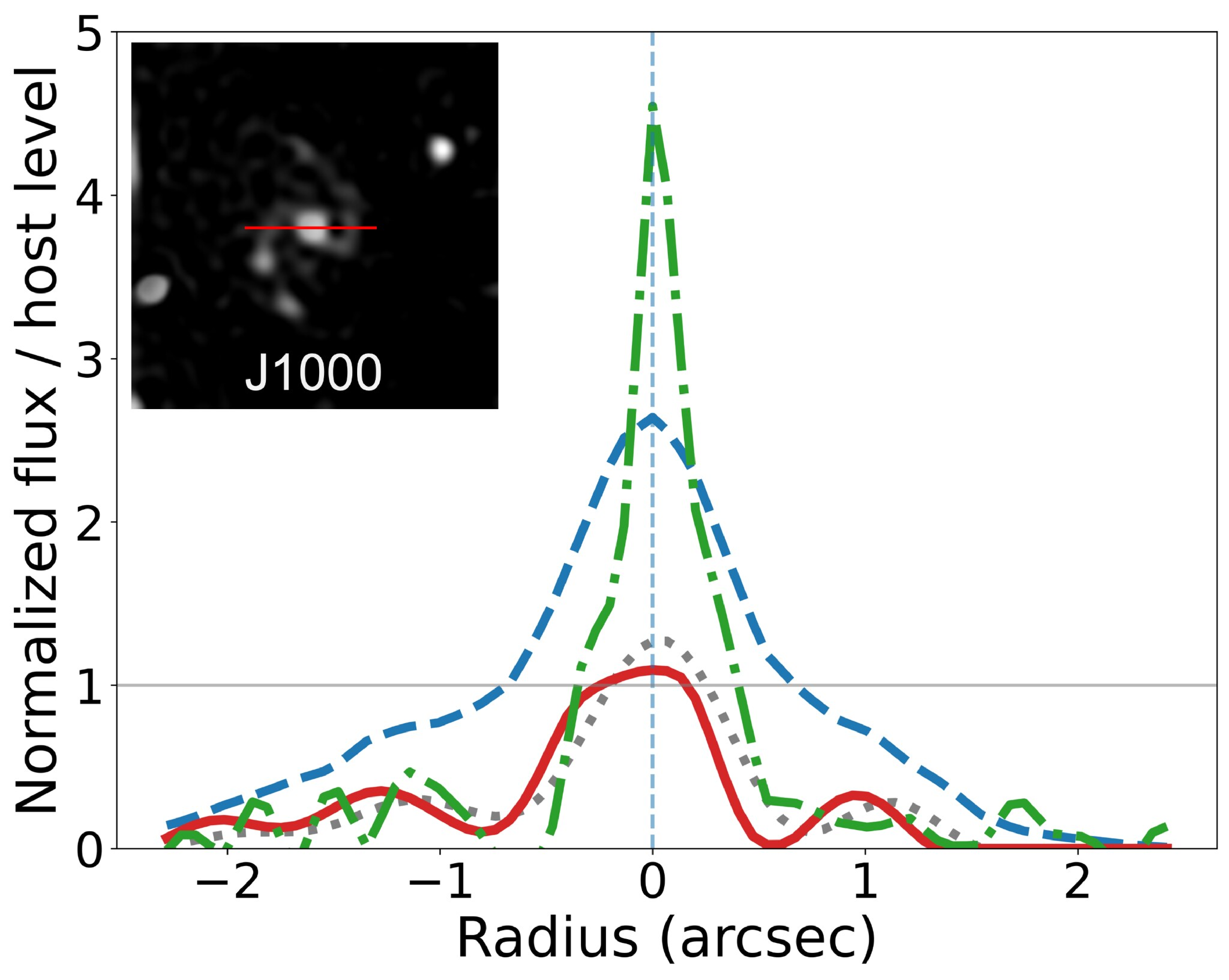} 
  \includegraphics[width=7cm]{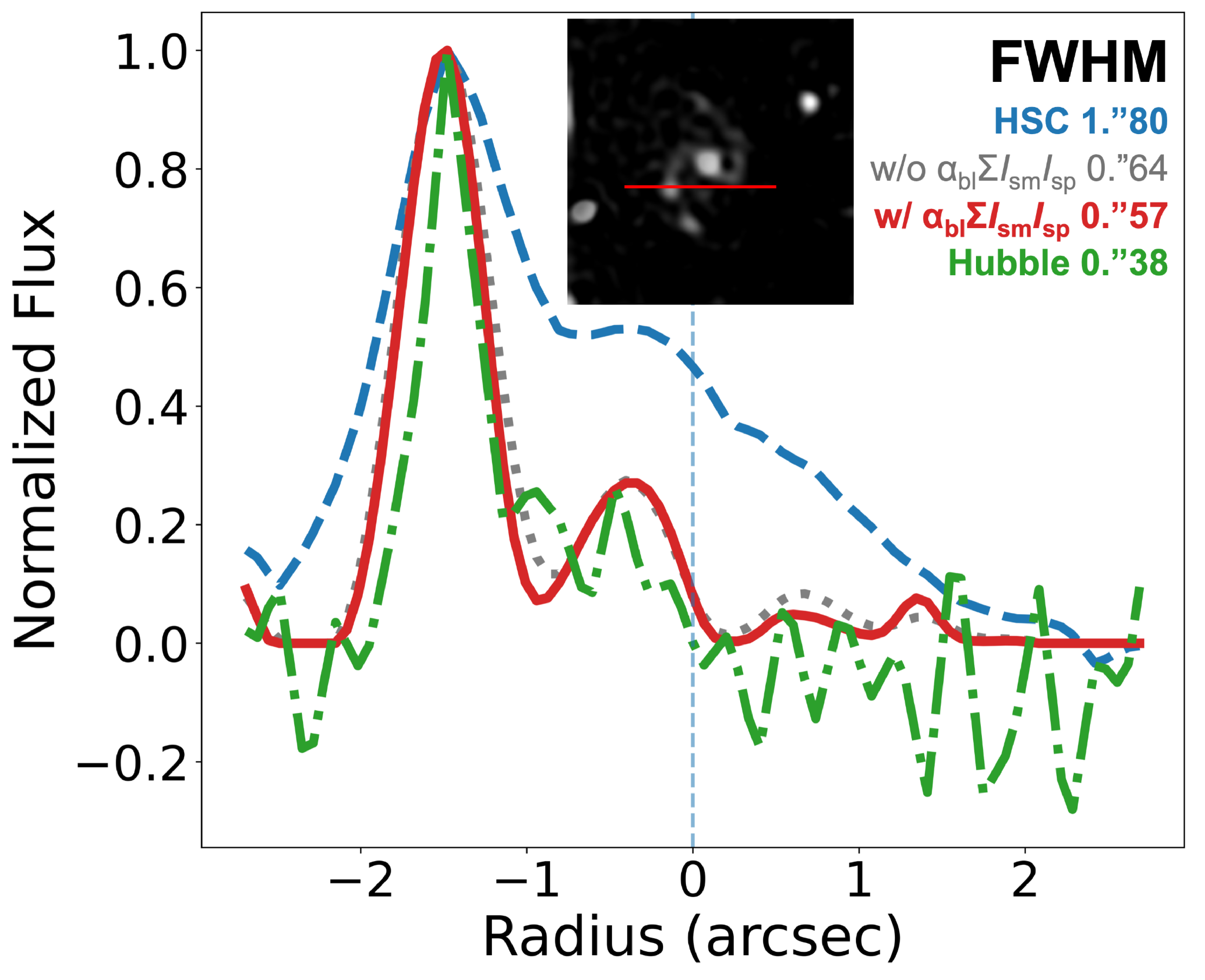} \\
 \end{center}
 \caption{Comparison of the flux profiles. The left panels show the flux profiles along the $x$-axis passing through the image centers, while the right panels show the profiles passing through the spiral arms. Note that the spiral arm profile is not defined for the artificial AGN because the host galaxy is constructed using a simple S\'ersic profile. The red solid line represents the results obtained with the point-source balance constraint, and the gray dotted line shows the results without this constraint. The blue dashed line corresponds to the HSC image, and the green dash-dotted line indicates the Hubble image. The vertical line indicates the image center. The horizontal line in the left panels represents the flux level near the peak of the host galaxy. The inset images display the smooth components used for the profile extraction, where the red horizontal lines indicate the measurement positions. To extract the flux profiles, pixel values are summed over $\Delta y = \pm 1\mbox{--}2$ pixels in the $y$-direction. For the central profiles, the vertical axis is normalized by a representative host-galaxy flux level $f_{\rm host}$, so that the plotted quantity is $f/f_{\rm host}$, where $f$ is the measured flux profile. The value of $f_{\rm host}$ is estimated by selecting a bright axis through the galaxy center and evaluating a region located approximately $1$--$2$ PSF FWHM from the center along that axis. The reference level $f_{\rm host}$ is defined as the mean of the brightest 10\% of pixel values within this region, which minimizes contamination from the central point source while providing a representative flux scale. For the spiral-arm profiles, this normalization is not applied; instead, each profile is normalized by its maximum value.}
 \label{fig:profiles}
\end{figure*}

This section presents the results of applying the deconvolution method to both artificial and real AGNs. As shown below, qualitative and quantitative evaluations demonstrate that the spatial resolution of HSC is improved to a level comparable to that of Hubble and that the AGN point-source component is effectively removed.

First, we qualitatively examine the reconstructed images to evaluate the performance of the spatial resolution enhancement and the point-source removal. Figure~\ref{fig:images_agn} shows the reconstructed images. As shown in the smooth component $\Ism$ (figure~\ref{fig:images_agn}b) and the sparse component $\Isp$ (figure~\ref{fig:images_agn}c), both the artificial AGN and the real AGNs are well separated into the host galaxies and the central point sources. No significant over-subtraction or residual flux is shown at the central region of the host galaxies. The spatial resolution of the smooth images $\Ism$ is improved relative to the HSC observed images (figure~\ref{fig:images_agn}a). In the real AGNs, host-galaxy structures such as spiral arms and clumpy substructures are clearly visible with high contrast and sharpness. The sum of the smooth and sparse components ($\Ism+\Isp$) closely resembles the reference Hubble images (figure~\ref{fig:images_agn}e). Some host-galaxy structures which are not clearly identified in the Hubble images are recovered with high contrast in the reconstructed images $\Ism + \Isp$, owing to the greater depth of the HSC-SSP data than that of the Hubble COSMOS-Wide field data. We compare the reconstructed $\Ism$ image of the artificial AGN with the model host galaxy in figure~\ref{fig:images_host}. The host galaxy of the artificial AGN is generated using the S\'ersic profile, allowing for an accurate image comparison. Figure~\ref{fig:images_host} shows that the central point source is effectively removed in the reconstructed image. 

Next, we quantitatively evaluate the performance of the reconstructed images. In figure \ref{fig:profiles}, we show the flux profiles at the galaxy center (left) and along the spiral arms (right), which are used to assess the effectiveness of the point-source removal and the spatial resolution enhancement, respectively. In the central profiles (left panels), the result obtained with our method (red solid line) shows that the central point source is effectively removed compared to both the HSC and Hubble images, with the central flux reduced to the level of the host galaxy. In the spiral-arm profiles (right panels), the reconstructed images with our method (red solid line) exhibit a clear improvement in spatial resolution, approaching that of the Hubble images. The FWHM values for the HSC, reconstructed, and Hubble images are $1\farcs06$, $0\farcs57$, and $0\farcs55$ for J0959, and $1\farcs80$, $0\farcs57$, and $0\farcs38$ for J1000, respectively. The FWHM values obtained with our method are substantially smaller than those of the HSC images—indicating roughly a factor of $\sim2-3$ improvement—and, for J0959, become comparable to that of the Hubble images.

Finally, we examine the effect of the point-source balance constraint $\alphabl$. The gray dotted lines in figure~\ref{fig:profiles} show the flux profiles obtained without the $\alphabl$ term. For the artificial AGN, there is no significant difference in the point-source removal performance between the cases with and without $\alphabl$. In contrast, for the real AGNs, the inclusion of the $\alphabl$ term improves both the spatial resolution and the point-source removal performance. The $\alphabl$ term reduces the FWHM values of the spiral-arm profiles to $0\farcs57$ for both J0959 and J1000, compared with $0\farcs67$ for J0959 and $0\farcs64$ for J1000 without $\alphabl$. These results indicate that the point-source balance constraint $\alphabl$ is effective for both spatial resolution enhancement and point-source removal.

\section{Conclusions}\label{sec:conclusions}

We have developed a new PSF deconvolution algorithm that simultaneously enhances the spatial resolution of host galaxies and removes the bright central point source in AGN images. The intrinsic image is reconstructed by decomposing the input into two components: an extended component $\Ism$ and a point-source component $\Isp$ (figure~\ref{fig:overview}). The reconstruction incorporates the three constraints --- smooth, sparse, and point-source balance (figure~\ref{fig:hyper3d}). We apply the method to both artificial and $z \sim 0-1$ real AGNs observed with HSC, and evaluate its performance both qualitatively and quantitatively. The reconstructed images demonstrate improved spatial resolution, approaching that of Hubble, and successful point-source removal.

The main results are as follows:

\begin{itemize}

\item We assess the stability and accuracy of the reconstruction using the Hellinger distance and RMSE. The Hellinger distance rapidly decreases at the early stage of the reconstruction and remains low and stable thereafter (figure~\ref{fig:hellinger}). The RMSE values are small for both the HSC-convolved reconstruction and the comparison with the Hubble images, indicating high reconstruction accuracy (table~\ref{tab:rmse}).

\item We evaluate the point-source removal performance using both the decomposed images and the flux profiles. The extended component $\Ism$ shows that the central point source is removed without significant over- or under-subtraction (figures~\ref{fig:images_agn}b and \ref{fig:images_host}), while the sparse component $\Isp$ concentrates the point-source flux at the image center (figure~\ref{fig:images_agn}c). The central flux profile is consistent with the host-galaxy level ($f / f_{\mathrm{host}} \sim 1$; figure~\ref{fig:profiles}, left). For the real AGNs, the point-source removal performance is improved compared to the case without the point-source balance constraint (the red solid and the gray dotted lines in figure~\ref{fig:profiles}, left).

\item We evaluate the spatial resolution enhancement using both the reconstructed images and the flux profiles. The smooth component $\Ism$ reveals fine host-galaxy structures such as spiral arms with high contrast (figure~\ref{fig:images_agn}b), and the combined images ($\Ism + \Isp$) closely match the Hubble images (figure~\ref{fig:images_agn}e). The spiral-arm profiles show that the reconstructed FWHM is comparable to that of Hubble (figure~\ref{fig:profiles}, right), indicating a resolution improvement by a factor of $\sim2-3$.

\end{itemize}

These results demonstrate that our algorithm is effective for studying AGN host-galaxy morphologies. This deconvolution algorithm is also applicable to wide-field surveys such as Vera C. Rubin Observatory, the Euclid Space Telescope, and the Roman Space Telescope, enabling statistical studies of large AGN samples. The method can be further extended to studies of SMBH--bulge relations \citep{1998AJ....115.2285M, 2000ApJ...539L...9F} and to the extraction of other structures such as giant stellar clumps \citep{2015ApJ...800...39G, 2016ApJ...821...72S}. 


\begin{ack}
We thank Mikio Morii for his significant contributions to the early development of this work. We are also grateful to Hiroshi Akitaya, Fumi Egusa, Daichi Kashino, Kazumi Murata, Masafusa Onoue, Yuma Sugahara, Takumi S. Tanaka, and Masafumi Yagi for providing helpful comments and suggestions.  This work was supported by the JSPS KAKENHI Grant Number 25K00223.
\end{ack}

\section*{Data availability} 
To ensure reproducibility and to facilitate further AGN studies, we will make the code for the algorithm developed in this work publicly available upon acceptance of the manuscript.


\section*{Appendix 1. Effect of the Pixel Interpolation on the Reconstruction Results}\label{sec:subpixel}

\begin{figure*}
 \begin{center}
  \includegraphics[width=16cm]{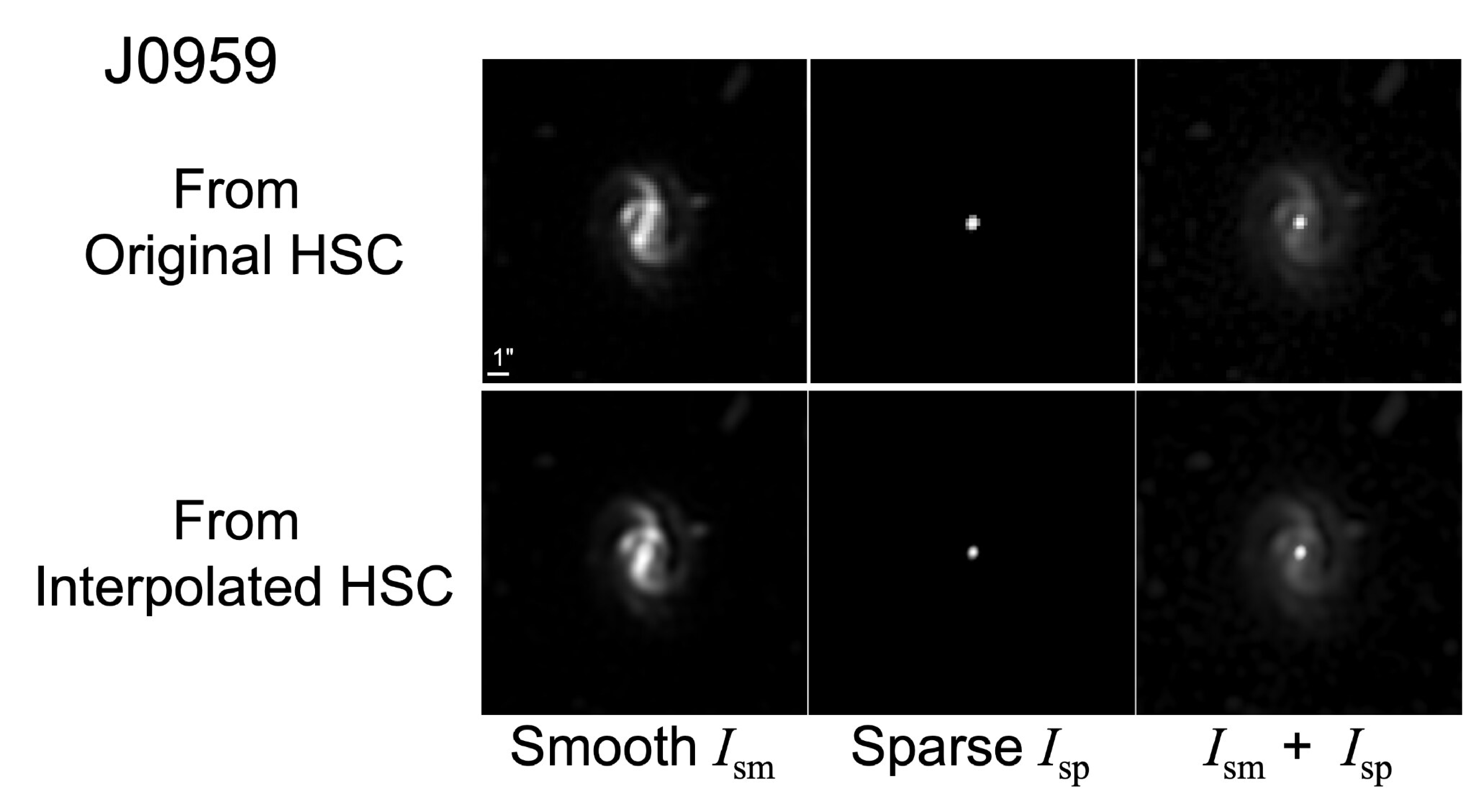}
  \includegraphics[width=16cm]{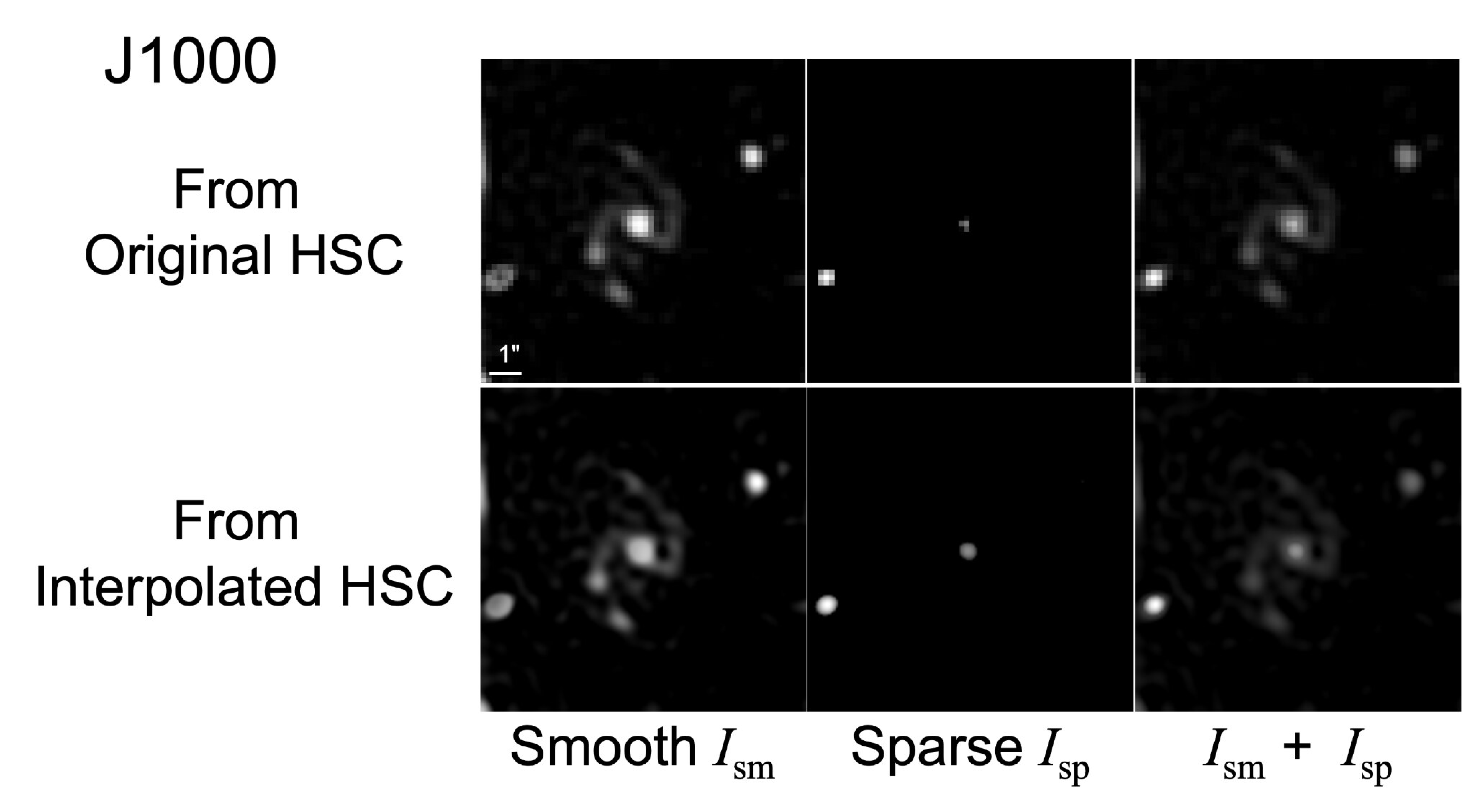} 
 \end{center}
\caption{Comparison of the reconstructed images obtained from the original pixel-size and the linearly interpolated input images for the real AGNs J0959 ($z = 0.34$) and J1000 ($z = 1.38$). The top and bottom rows for each object show the results for the original-size and the interpolated images, respectively. In each row, the panels display the smooth component $\Ism$, the sparse component $\Isp$, and the sum of the smooth and sparse components ($\Ism + \Isp$) from left to right. The white solid line in the upper-left panel of each object indicates a scale of $1^{\prime\prime}$. No significant differences are found between the results obtained from the original-size and the interpolated images, indicating that the linear interpolation used in this work to match the image size has a negligible effect on the reconstruction results.}
\label{fig:interpolated}
\end{figure*}

To assess the effect of linear pixel interpolation, we compare the reconstructed images obtained from the original images with those  obtained from the interpolated inputs in figure~\ref{fig:interpolated}. There is no significant difference in the structure of the central target objects. This comparison indicates that the linear interpolation does not strongly affect the reconstructed images.

\section*{Appendix 2. Results of the Hyperparameter Grid Searches}\label{sec:grid}

\begin{figure*}
 \begin{center}
  \includegraphics[width=16cm]{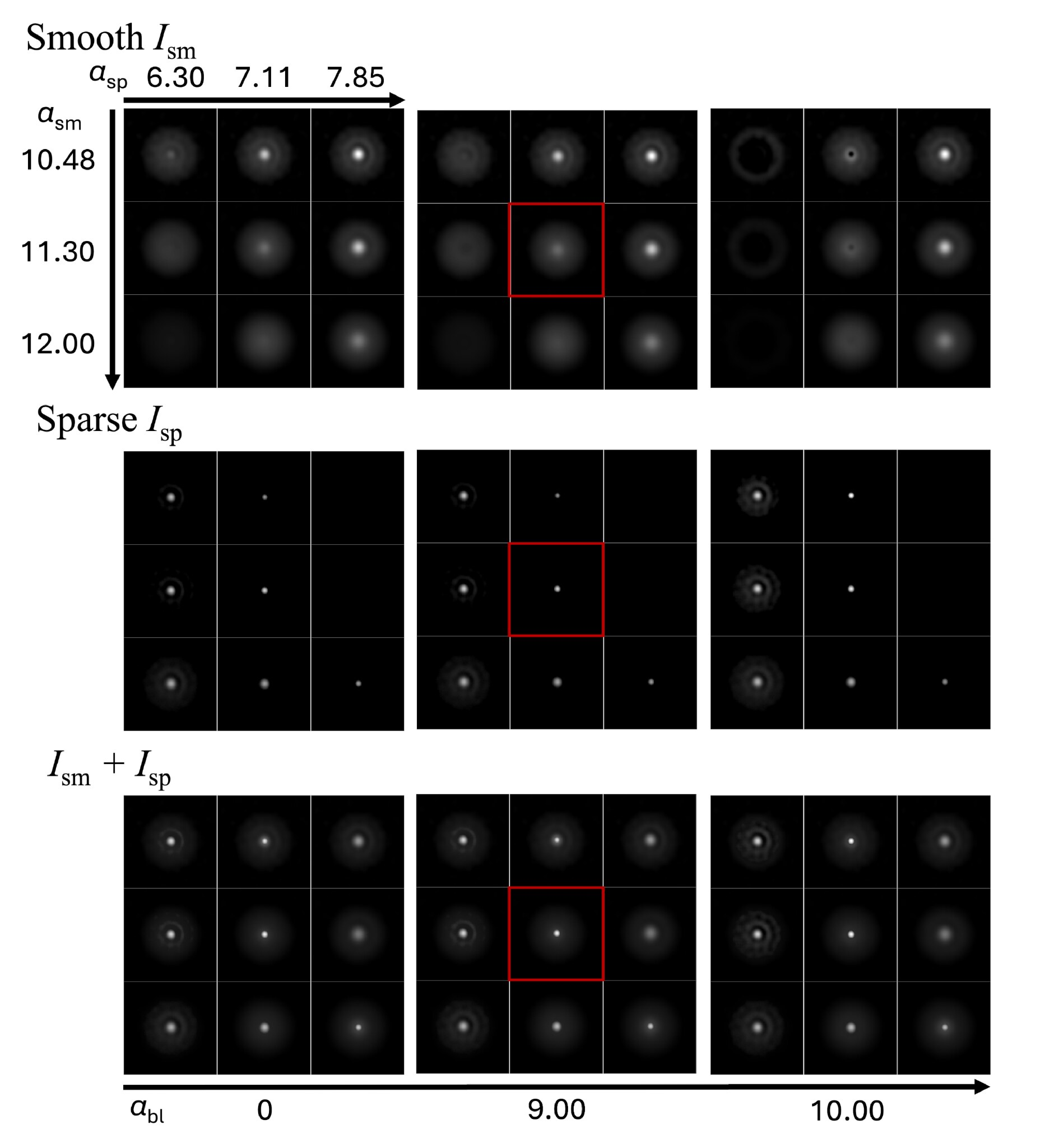}
 \end{center}
\caption{The grid-search results for the artificial AGN. The top, middle, and bottom panels display the smooth component $\Ism$, the sparse component $\Isp$, and the combined reconstructed image $\Ism + \Isp$, respectively. Within each $\alphabl$ block, the columns from left to right correspond to increasing $\alphasp$, while the rows from top to bottom correspond to increasing $\alphasm$. The numbers shown in the figure indicate the values of $(\log_{10}\alphasp, \log_{10}\alphasm, \log_{10}\alphabl)$, except for the cases where $\alphabl=0$.}
\label{fig:grid_artificial}
\end{figure*}

\begin{figure*}
 \begin{center}
  \includegraphics[width=16cm]{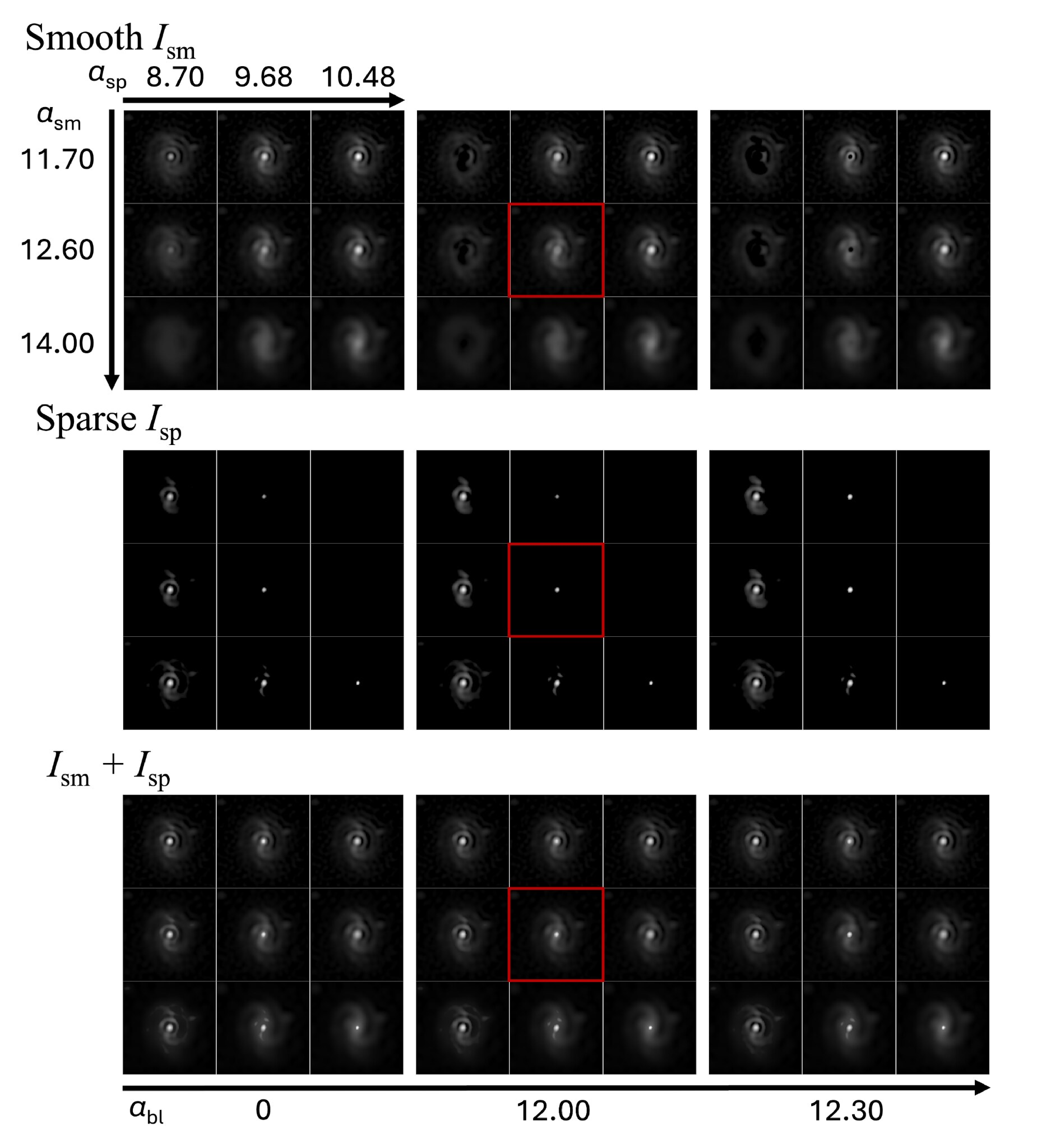}
 \end{center}
\caption{Same as figure \ref{fig:grid_artificial}, but for J0959.}
\label{fig:grid_j0959}
\end{figure*}

\begin{figure*}
 \begin{center}
  \includegraphics[width=16cm]{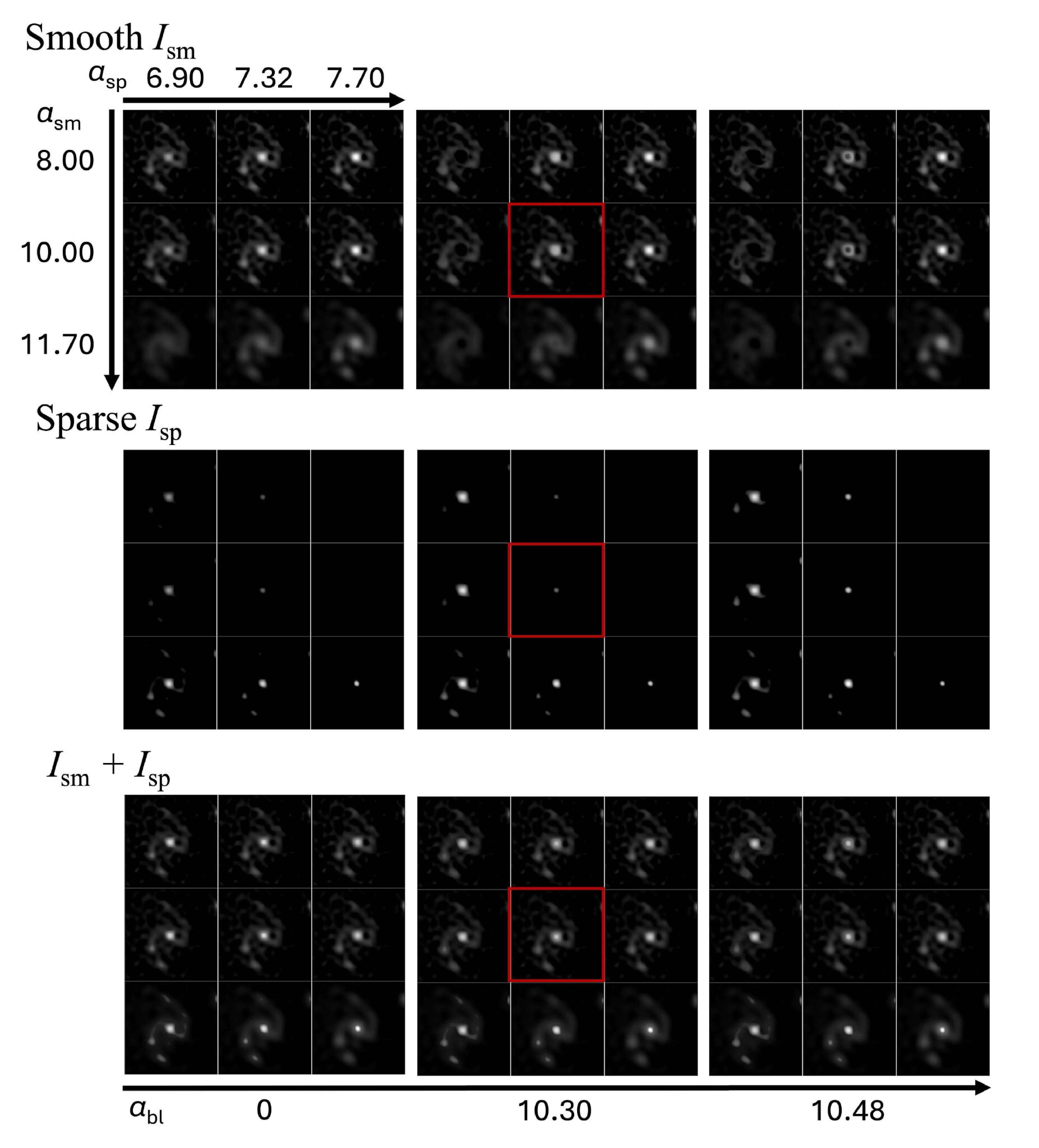}
 \end{center}
\caption{Same as figure \ref{fig:grid_artificial}, but for J1000.}
\label{fig:grid_j1000}
\end{figure*}

We present the results of the hyperparameter grid searches in figures \ref{fig:grid_artificial}, \ref{fig:grid_j0959}, and \ref{fig:grid_j1000}. 

\clearpage

\clearpage

\clearpage

\begin{thebibliography}{}

\bibitem[Aihara et al.(2022)]{2022PASJ...74..247A}
Aihara, H., et al. 2022, \pasj, 74, 247

\bibitem[Beck \& Teboulle(2009)]{articles}
Beck, A., \& Teboulle, M. 2009, SIAM J. Imaging Sci., 2, 183

\bibitem[Bonaventura et al.(2025)]{2025ApJ...978...74B}
Bonaventura, N., et al. 2025, \apj, 978, 74

\bibitem[Bonaventura et al.(2026)]{2026ApJ...997...47B}
Bonaventura, N., et al. 2026, \apj, 997, 47

\bibitem[Cisternas et al.(2011)]{2011ApJ...726...57C}
Cisternas, M., et al. 2011, \apj, 726, 57

\bibitem[Chen et al.(2023)]{2023Natur.616...45C}
Chen, Y.-C., et al. 2023, \nat, 616, 45

\bibitem[Conselice(2003)]{2003ApJS..147....1C}
Conselice, C. J. 2003, \apjs, 147, 1

\bibitem[Daubechies et al.(2004)]{article}
Daubechies, I., et al. 2004, CPAM, 57, 1413

\bibitem[Decarli et al.(2019)]{2019ApJ...880..157D}
Decarli, R., et al. 2019, \apj, 880, 157

\bibitem[Decarli et al.(2024)]{2024A&A...689A.219D}
Decarli, R., et al. 2024, \aap, 689, A219

\bibitem[Di Matteo et al.(2005)]{2005Natur.433..604D}
Di Matteo, T., et al. 2005, \nat, 433, 604

\bibitem[Ding et al.(2022)]{2022ApJ...939L..28D}
Ding, X., et al. 2022, \apjl, 939, L28

\bibitem[Ding et al.(2025)]{2025ApJ...993...91D}
Ding, X., et al. 2025, \apj, 993, 91

\bibitem[Donley et al.(2018)]{2018ApJ...853...63D}
Donley, J. L., et al. 2018, \apj, 853, 63

\bibitem[Farrens et al.(2017)]{2017A&A...601A..66F}
Farrens, S., et al. 2017, \aap, 601, A66

\bibitem[Ferrarese \& Merritt(2000)]{2000ApJ...539L...9F}
Ferrarese, L., \& Merritt, D. 2000, \apjl, 539, L9

\bibitem[Gabor et al.(2009)]{2009ApJ...691..705G}
Gabor, J. M., et al. 2009, \apj, 691, 705

\bibitem[Gan et al.(2021)]{2021arXiv210309711G}
Gan, F. K., et al. 2021, arXiv e-prints, arXiv:2103.09711

\bibitem[Getachew-Woreta et al.(2022)]{2022MNRAS.514..607G}
Getachew-Woreta, T., et al. 2022, \mnras, 514, 607

\bibitem[Glikman et al.(2015)]{2015ApJ...806..218G}
Glikman, E., et al. 2015, \apj, 806, 218

\bibitem[Goulding et al.(2018)]{2018PASJ...70S..37G}
Goulding, A. D., et al. 2018, \pasj, 70, S37

\bibitem[Guo et al.(2015)]{2015ApJ...800...39G}
Guo, Y., et al. 2015, \apj, 800, 39

\bibitem[Heckman \& Best(2014)]{2014ARA&A..52..589H}
Heckman, T. M., \& Best, P. N. 2014, \araa, 52, 589

\bibitem[Hopkins et al.(2005)]{2005ApJ...630..705H}
Hopkins, P. F., et al. 2005, \apj, 630, 705

\bibitem[Hopkins et al.(2006)]{2006ApJS..163....1H}
Hopkins, P. F., et al. 2006, \apjs, 163, 1

\bibitem[Kahn et al.(2026)]{2026arXiv260420195K}
Kahn, S., et al. 2026, arXiv e-prints, arXiv:2604.20195

\bibitem[Kartaltepe et al.(2015)]{2015ApJS..221...11K}
Kartaltepe, J. S., et al. 2015, \apjs, 221, 11

\bibitem[Kocevski et al.(2012)]{2012ApJ...744..148K}
Kocevski, D. D., et al. 2012, \apj, 744, 148

\bibitem[Koekemoer et al.(2007)]{2007ApJS..172..196K}
Koekemoer, A. M., et al. 2007, \apjs, 172, 196

\bibitem[Kormendy \& Ho(2013)]{2013ARA&A..51..511K}
Kormendy, J., \& Ho, L. C. 2013, \araa, 51, 511

\bibitem[La Marca et al.(2026)]{2026A&A...708A.373L}
La Marca, A., et al. 2026, \aap, 708, A373

\bibitem[Lambrides et al.(2021)]{2021ApJ...919..129L}
Lambrides, E. L., et al. 2021, \apj, 919, 129

\bibitem[Leist et al.(2024)]{2024AJ....167...96L}
Leist, M. T., et al. 2024, \aj, 167, 96

\bibitem[Lotz et al.(2004)]{2004AJ....128..163L}
Lotz, J. M., et al. 2004, \aj, 128, 163

\bibitem[Lucy(1974)]{1974AJ.....79..745L}
Lucy, L. B. 1974, \aj, 79, 745

\bibitem[Magain et al.(1998)]{1998ApJ...494..472M}
Magain, P., et al. 1998, \apj, 494, 472

\bibitem[Magorrian et al.(1998)]{1998AJ....115.2285M}
Magorrian, J., et al. 1998, \aj, 115, 2285

\bibitem[Marshall et al.(2025)]{2025A&A...702A.174M}
Marshall, M. A., et al. 2025, \aap, 702, A174

\bibitem[Matsuoka et al.(2018)]{2018ApJ...869..150M}
Matsuoka, Y., et al. 2018, \apj, 869, 150

\bibitem[Mechtley et al.(2016)]{2016ApJ...830..156M}
Mechtley, M., et al. 2016, \apj, 830, 156

\bibitem[Millon et al.(2024)]{2024AJ....168...55M}
Millon, M., et al. 2024, \aj, 168, 55

\bibitem[Miyazaki et al.(2018)]{2018PASJ...70S...1M}
Miyazaki, S., et al. 2018, \pasj, 70, S1

\bibitem[Morii et al.(2019)]{2019PASJ...71...24M}
Morii, M., et al. 2019, \pasj, 71, 24

\bibitem[Morii et al.(2024)]{2024PASJ...76..272M}
Morii, M., et al. 2024, \pasj, 76, 272

\bibitem[Murata \& Takeuchi(2022)]{2022PASJ...74.1329M}
Murata, K., \& Takeuchi, T. T. 2022, \pasj, 74, 1329

\bibitem[Onoue et al.(2025)]{2025NatAs...9.1541O}
Onoue, M., et al. 2025, Nature Astronomy, 9, 1541

\bibitem[Peng et al.(2002)]{2002AJ....124..266P}
Peng, C. Y., et al. 2002, \aj, 124, 266

\bibitem[Peng et al.(2010)]{2010AJ....139.2097P}
Peng, C. Y., et al. 2010, \aj, 139, 2097

\bibitem[Puetter et al.(2005)]{2005ARA&A..43..139P}
Puetter, R. C., et al. 2005, \araa, 43, 139

\bibitem[Richardson(1972)]{1972JOSA...62...55R}
Richardson, W. H. 1972, J. Opt. Soc. Am., 62, 55

\bibitem[Schawinski et al.(2017)]{2017MNRAS.467L.110S}
Schawinski, K., et al. 2017, \mnras, 467, L110

\bibitem[S\'ersic(1963)]{1963BAAA....6...41S}
S\'ersic, J. L. 1963, BAAA, 6, 41

\bibitem[Shibuya et al.(2016)]{2016ApJ...821...72S}
Shibuya, T., et al. 2016, \apj, 821, 72

\bibitem[Shibuya et al.(2025)]{2025PASJ...77...21S}
Shibuya, T., et al. 2025, \pasj, 77, 21

\bibitem[Simmons \& Urry(2008)]{2008ApJ...683..644S}
Simmons, B. D., \& Urry, C. M. 2008, \apj, 683, 644

\bibitem[Tanaka et al.(2025)]{2025ApJ...995...21T}
Tanaka, T. S., et al. 2025, \apj, 995, 21

\bibitem[Tibshirani(1996)]{10.1111/j.2517-6161.1996.tb02080.x}
Tibshirani, R. 1996, JRSSB, 58, 267

\bibitem[Tikhonov(1963)]{Tikhonov:1963}
Tikhonov, A. N. 1963, Soviet Math. Dokl., 4, 1035

\bibitem[Vijarnwannaluk et al.(2025)]{2025ApJ...994..265V}
Vijarnwannaluk, B., et al. 2025, \apj, 994, 265

\bibitem[Villforth et al.(2017)]{2017MNRAS.466..812V}
Villforth, C., et al. 2017, \mnras, 466, 812

\bibitem[Weaver et al.(2022)]{2022ApJS..258...11W}
Weaver, J. R., et al. 2022, \apjs, 258, 11

\bibitem[Wylezalek et al.(2022)]{2022ApJ...940L...7W}
Wylezalek, D., et al. 2022, \apjl, 940, L7

\bibitem[Zakamska et al.(2019)]{2019MNRAS.489..497Z}
Zakamska, N. L., et al. 2019, \mnras, 489, 497

\end{thebibliography}
\end{document}